\documentclass{article}
\usepackage{graphicx}  % standard LaTeX graphics tool;
\usepackage{amsmath}   % For getting proper math eqns
\usepackage[compress]{cite}
\usepackage{amssymb}   % Used for getting various symbols eg. gtrsim, lesssim etc.
\usepackage{bm} % For bold math (esp of lower greek letters)
\usepackage{dcolumn}% Align table columns on decimal point
\usepackage{color}
\usepackage{mathrsfs}
\usepackage{enumitem}
\usepackage{amsfonts}
\usepackage[caption=false]{subfig}
\usepackage{varioref}
\usepackage{textcomp}
\usepackage{tikz}
\usepackage{tikz,pgfplots}
\usepackage{pst-plot}
\usepackage{gensymb}
\RequirePackage[colorlinks,citecolor=blue,urlcolor=magenta,linkcolor=blue]{hyperref}
\input epsf
\def\gr{general relativity}

\labelformat{section}{Section #1} 
\labelformat{subsection}{Section #1} 
\labelformat{subsubsection}{Section #1}
\labelformat{subsubsubsection}{Section #1}
\labelformat{equation}{Eq.~(#1)} 
\labelformat{figure}{Fig.~#1} 
\labelformat{subfigure}{Fig.~\thefigure#1} 
\labelformat{table}{Tab.~#1} 
\labelformat{appendix}{Appendix #1}
\title{Do shadows of Sgr A* and M87* indicate black holes with a magnetic monopole charge?  }
\author{Indrani Banerjee\footnote{banerjeein@nitrkl.ac.in}~$^{1}$, Subhadip Sau\footnote{subhadipsau2@gmail.com}~$^{2,3}$ and Soumitra SenGupta\footnote{tpssg@iacs.res.in}~$^{3}$\\
{$^{1}$\small{Department of Physics and Astronomy, National Institute of Technology, Rourkela}}\\
{$^{2}$\small{Department of Physics, Jhargram Raj College, Jhargram, West Bengal-721507}}\\
{$^{3}$\small{School of Physical Sciences, Indian Association for the Cultivation of Science, Kolkata-700032}}}
\date{ }  %% This command  will suppress printing the date. 
\begin{document} 
\maketitle
%%%%%%%%%%%%%%%%%%%%%%%%%%%%%%%%%%%%%%%%%%%%%%%%%%%%%%%%%%%%%%%%%%%%%%%%%%%%%%%%%%%%%%%%%%%%%%%%%%%
%%%%%%%%%%%%%%%%%%%%%%%%%%%%%%%%%%%%%%%%%%%%%%%%%%%%%%%%%%%%%%%%%%%%%%%%%%%%%%%%%%%%%%%%%%%%%%%%%%%
%%%%%%%%%%%%%%%%%%%%%%%%%%%%%%%%%%%%%%%%%%%%%%%%%%%%%%%%%%%%%%%%%%%%%%%%%%%%%%%%%%%%%%%%%%%%%%%%%%%
\begin{abstract}
We study the prospect of Bardeen black holes in explaining the observed shadow of Sgr A* and M87*. Bardeen black holes are regular black holes endowed with a magnetic monopole charge that arise in Einstein gravity coupled to non-linear electrodynamics. These black holes are interesting as they can evade the $r=0$ curvature singularity arising in \gr. It is therefore worthwhile to look for signatures of Bardeen black holes in astrophysical observations. With two successive release of black hole images by the Event Horizon Telescope (EHT) collaboration, the scope to test the nature of strong gravity has substantially increased. We compare the theoretically computed shadow observables with the observed image of Sgr A* and M87*. Our analysis reveals that while the observed angular diameter of M87* favors the Kerr scenario, the shadow of Sgr A* can be better explained by the Bardeen background. This indicates that although rare, certain black holes exhibit a preference towards regular black holes like the Bardeen spacetime.
\end{abstract}

%%%%%%%%%%%%%%%%%%%%%%%%%%%%%%%%%%%%%%%%%%%%%%%%%%%%%%%%%%%%%%%%%%%%%%%%%%%%%%%%%%%%%%%%%%%%%%%%%%%
%%%%%%%%%%%%%%%%%%%%%%%%%%%%%%%%%%%%%%%%%%%%%%%%%%%%%%%%%%%%%%%%%%%%%%%%%%%%%%%%%%%%%%%%%%%%%%%%%%%
%%%%%%%%%%%%%%%%%%%%%%%%%%%%%%%%%%%%%%%%%%%%%%%%%%%%%%%%%%%%%%%%%%%%%%%%%%%%%%%%%%%%%%%%%%%%%%%%%%%
%%%%%%%%%%%%%%%%%%%%%%%%%%%%%%%%%%%%%%%%%%%%%%%%%%%%%%%%%%%%%%%%%%%%%%%%%%%%%%%%%%%%%%%%%%%%%%%%%%%
%%%%%%%%%%%%%%%%%%%%%%%%%%%%%%%%%%%%%%%%%%%%%%%%%%%%%%%%%%%%%%%%%%%%%%%%%%%%%%%%%%%%%%%%%%%%%%%%%%%
%%%%%%%%%%%%%%%%%%%%%%%%%%%%%%%%%%%%%%%%%%%%%%%%%%%%%%%%%%%%%%%%%%%%%%%%%%%%%%%%%%%%%%%%%%%%%%%%%%%
\section{Introduction}
Black holes which mark a one-way passage to the world of the unknown are the most enigmatic predictions of general relativity. These objects are surrounded by an event horizon from where nothing can escape to the outside world. The horizon represents the characteristic radius within which if a mass distriution is compressed, no signals from it can reach the external observer. The existence of black holes have received ample observational confirmation since the last fifty years with the launch of X-ray satellites like RXTE, Chandra, etc. However, these are indirect detections of black holes. With two successive ground breaking discoveries, namely, the gravitational waves from the merger of binary compact objects \cite{Abbott:2016blz,TheLIGOScientific:2017qsa} and the image of the black hole \cite{Fish:2016jil,Akiyama:2019cqa,Akiyama:2019brx,Akiyama:2019sww,Akiyama:2019bqs,Akiyama:2019fyp,Akiyama:2019eap}, direct detection of these esoteric objects have become a reality. Yet the discovery of black holes mark at the same time the greatest feat and fiasco of general relativity. This is because these objects have a curvature singularity at the origin where the physical observables diverge and the theory loses all its predictability. The celebrated theorems of Hawking and Penrose state that black holes in \gr\ eventually end up in singularities \cite{Hawking:1973uf,Penrose:1964wq,Hawking:1976ra,Christodoulou:1991yfa}. It is believed that a suitable quantum theory of gravity may cure these problems and various quantum gravity models have been proposed in this regard \cite{Horava:1995qa,Horava:1996ma,Polchinski:1998rq,Polchinski:1998rr,Ashtekar:2006rx,Kothawala:2013maa}. 

However, there exist a classical route to address the singularity issue in GR which involves investigating black holes in Einstein gravity coupled to non-linear electrodynamics \cite{Ayon-Beato:2000mjt,Ayon-Beato:1998hmi,Ayon-Beato:2004ywd,Bronnikov:2000vy,Borde:1996df,Barrabes:1995nk,Ayon-Beato:1999qin,Bonanno:2000ep,Nicolini:2005vd,Myung:2007qt,PhysRevLett.96.031103}. These models arise in the low energy limit of various quantum gravity theories where quantum fluctuations during the last stages of a stellar collapse can potentially avoid the unbounded increase in the spacetime curvature thereby giving rise to regular cores. For these black holes the metric and the curvature invariants assume finite values at all spacetime points. The regular core can be either de Sitter or Minkowski type \cite{Frolov:1988vj,Mukhanov:1991zn,Brandenberger:1993ef,Ansoldi:2008jw,1966JETP...22..378G,Fan:2016hvf}. In this work we shall consider regular black holes which are endowed with a repulsive de Sitter core. Inspired by the ideas of Sakharov \cite{1966JETP...22..241S} and Gliner \cite{1966JETP...22..378G} such black holes were first proposed by Bardeen as solution of Einstein equations coupled to non-linear electrodynamics. Further investigations by Ayon-Beato and Garcia \cite{Ayon-Beato:2000mjt,Ayon-Beato:1998hmi,Ayon-Beato:2004ywd,Ayon-Beato:1999qin} revealed that the Bardeen black holes are associated with the gravitational field of a non-linear magnetic monopole of a self-gravitating magnetic field. Since then there has been a lot of work in the direction of black holes in non-linear electrodynamics \cite{Ansoldi:2008jw,1966JETP...22..378G,Fan:2016hvf,Burinskii:2002pz,Berej:2006cc,Balart:2014cga,Junior:2015fya,Sajadi:2017glu} and several other regular black hole solutions have been put forward \cite{Ayon-Beato:1999qin,Ayon-Beato:2000mjt,Dymnikova:1992ux,Bronnikov:2000vy,Ansoldi:2008jw,Bronnikov:2017tnz,Bronnikov:2005gm,Kumar:2020ltt,Ghosh:2018bxg,PhysRevLett.96.031103,Toshmatov:2018cks}. Since astrophysical black holes are generally rotating, we shall consider the axisymmetric counterpart of the Bardeen black holes in this work. The rotating solution is obtained by applying the Newman-Janis algorithm to the static spherically symmetric spacetime \cite{Newman:1965tw,Azreg-Ainou:2014pra,Azreg-Ainou:2014aqa,Azreg-Ainou:2014nra}.  

Observational signatures of regular black holes have been explored extensively \cite{Kumar:2020ltt,Kumar:2020yem,Allahyari:2019jqz,Vagnozzi:2022moj,Uniyal:2022vdu,Stuchlik:2014qja,Schee:2015nua,Schee:2016mjd,Stuchlik:2019uvf,Schee:2019gki,Banerjee:2022ffu,Banerjee:2021nza,Banerjee:2022chn}.
In this work we shall study the signatures of Bardeen black holes in the recently observed shadow of Sgr A* and M87*. The imaging of black holes by the Event Horizon Telescope collaboration has opened up a new and unique window to test the nature of gravity in the strong field regime. The shadow of a black hole represents the directions in the local sky from where light just escapes the event horizon and reaches the observer on earth \cite{Cunha:2018acu,Vries_1999,Gralla:2019xty,Abdujabbarov:2015xqa,Abdujabbarov:2016hnw}. As a result there is total absence of radiation from certain directions in the observer's sky epitomized by a dark patch which is known as the black hole shadow. Since the boundary of the shadow nearly traces the event horizon, the shape and size of the shadow bears the imprints of the background metric \cite{Gralla:2019xty,Bambi:2019tjh,Hioki:2009na,Vagnozzi:2019apd,Banerjee:2019nnj,Banerjee:2022iok}. As a result the black hole image turns out to be a useful observational tool to look for signatures of regular black holes, which is the primary goal of the present work. 

The paper is organized as follows: In \ref{SP4_S2} we have discussed the non-linear electrodynamics model which gives rise to Bardeen black holes. \ref{SP4_S3} includes derivation of the shadow outline for Bardeen black holes. \ref{SP4_S4} is dedicated in discussing the comparison of our theoretical findings with the observed shadow of M87* and Sgr A*. We conclude with a summary of our results with some scope for future work in \ref{S5}.

\section{Brief description of Bardeen black holes}\label{SP4_S2}
In this section we discuss the theoretical background related to Bardeen black holes, the first ever regular black hole proposed in the literature. Such black holes arise in Einstein gravity coupled to non-linear electrodynamics such that the associated action is given by \cite{Kumar:2020ltt,Ayon-Beato:2000mjt,Salazar:1987ap,Fan:2016hvf,Bronnikov:2017tnz,Toshmatov:2018cks},
\begin{flalign}
S=\dfrac{1}{16\pi G}\int d^{4}x \sqrt{-h}\left[R-4\mathcal{L}(F) \right]
\label{Eq1}
\end{flalign}
where $G$ is gravitational constant, $h$ is the determinant of metric tensor, $R$ is Ricci scalar, $\mathcal{L}$ is Lagrangian density for non-linear electrodynamics and the Faraday invariant is $F=F_{\alpha\beta}F^{\alpha\beta}$. Electromagnetic field strength tensor is $F_{\alpha\beta}=\partial_{\alpha}A_{\beta}-\partial_{\beta}A_{\alpha}$, where $A^{\alpha}$ is the associated vector potential.
The Lagrangian density corresponding to Bardeen black holes is given by \cite{Kumar:2020ltt},
\begin{align}
\mathcal{L}(F) = \frac{3\mathcal{M}}{|\tilde{g}|\tilde{g}^2}\bigg{[} \frac{\sqrt{2\tilde{g}^2 F}}{1 + \sqrt{2\tilde{g}^2 F}}\bigg{]}^{5/2}
\label{Eq2}
\end{align}
where $\tilde{g}$ is the magnetic monopole charge associated with a self-gravitating magnetic field. 
Variation of the action with respect to the metric yields the gravitational field equations,
\begin{align}
&\mathcal{G}^{\nu}_{\mu} = 2\bigg{[}\mathcal{L}_{F}({F}_{\mu\lambda}{F}^{\nu\lambda}) - \delta^{\nu}_{\mu}\mathcal{L}(F)\bigg{]}
\label{Eq3}
\end{align}
while variation with respect to the gauge field $A_\mu$ leads to the equation of motion for the electromagnetic field,
\begin{align}
\nabla_{\mu} (\mathcal{L}_{F}\mathcal{F}^{\beta\mu}) =0
\label{Eq4}
\end{align}
with $\mathcal{L}_{F} = \frac{\partial\mathcal{L}}{\partial F}$.
The static, spherically symmetric black hole solution of \ref{Eq3} assumes the form,
\begin{align}
ds^2=-\bigg(1-\frac{2m(r)}{r}\bigg)dt^2 + \bigg(1-\frac{2m(r)}{r}\bigg)^{-1}dr^2 +r^2 d\Omega^2
\label{Eq5}
\end{align}
where the mass function is denoted by $m(r)$. When the black hole possesses pure magnetic charge the solution of the gauge field is given by $dA=-gcos\theta d\phi$ while from \ref{Eq3} and \ref{Eq4} we note that the mass function satisfies the following two equations:
\begin{align}
\label{Eq6}
&-\frac{2m'(r)}{r^2} + 2 \mathcal{L}(F)=0 \nonumber \\
&-\frac{m''(r)}{r} + 2 \mathcal{L}(F) -\frac{2g^2}{r^4}\mathcal{L}_F=0
\end{align}
For the Lagrangian density given by \ref{Eq2}, solution of \ref{Eq6} gives rise to the following form of the mass function,
\begin{align}
m(r)=M\bigg \lbrace\frac{r^2}{r^2 + \tilde{g}^2}\bigg\rbrace^{3/2}
\label{Eq7}
\end{align}
where we denote the magnetic monopole charge parameter by $g=\tilde{g}^2$ in the rest of the paper.

Since astrophysical black holes are generally rotating one needs to find the rotating counterpart of \ref{Eq5}. This is accomplished by applying the Newman-Janis algorithm to the static, spherically symmetric seed metric \cite{Newman:1965tw,Azreg-Ainou:2014pra,Azreg-Ainou:2014aqa,Azreg-Ainou:2014nra}. The rotating counterpart of the Bardeen metric  in Boyer-Lindquist coordinates correspond to, 
\begin{flalign}
ds^{2}=& - \left(1-\dfrac{2m(r)r}{\Sigma}\right)dt^{2}+ \dfrac{\Sigma}{\Delta} dr^{2}+\Sigma ~d\theta^{2}  -  \dfrac{4m(r) a r \sin^{2}\theta }{\Sigma} dt d\phi \nonumber\\
& \hspace{5cm} +% \underbrace{\dfrac{\left[(r^{2}+a^{2})^{2}-\Delta_{KN}a^{2}\sin^{2}\theta\right]}{\Sigma_{KN}}}_
\left[ {(r^{2}+a^{2})+\frac{2 m(r) r}{\Sigma}a^{2}\sin^{2}\theta}\right] \sin^{2}\theta d\phi^{2}
\label{Eq8}
\end{flalign}
where
\begin{flalign}
\Delta & = r^{2}-2m(r) r +a^{2}\\
\Sigma & = r^{2}+a^{2}\cos^{2}\theta
\label{Eq9}
\end{flalign}
and the mass function $\lim_{r\to\infty}m(r)=M$ ($M$ is ADM mass of black hole) and $a$ is the spin parameter of the black hole. From \ref{Eq8} and in the rest of the paper, we use dimensionless quantities, e.g. $r\equiv r/M$, $a\equiv a/M$ and $g\equiv g/M^2$.

Since we are interested in black hole solutions, the horizon radius/radii must be real and positive. As the horizon comprises of a null surface with spherical topology, the corresponding radii are obtained by solving for the roots of $g^{rr}=\Delta=0$,
\begin{align}
r^2 + a^2 -2rm(r)=0
\end{align}
Requirement of a real, positive horizon imposes bounds on the magnitude of $g$: $0\lesssim g\lesssim 0.55$, the upper limit being obtained by putting $a=0$ \cite{Kumar:2018ple}. Moreover, for a given $g$ in the allowed range, it restricts the allowed values of $a$ \cite{Kumar:2018ple}. 
In the next section we derive the shape and size of the black hole shadow for the rotating Bardeen metric given by \ref{Eq8}. The theoretically obtained shadow is then compared with the observed image of Sgr A* and M87* to establish constraints on the metric parameters $g$ and $a$.

\section{Nature of shadow for Kerr-like solutions}\label{SP4_S3}
In this section we derive the nature of the shadow cast by Bardeen black holes. This requires investigating the geodesic motion of photons around black holes which may originate from a distant astrophysical object or from the accretion disk surrounding the black hole. When these photons arrive close to the event horizon a part of it gets trapped inside the horizon while the remaining photons escape to infinity. This creates a dark patch in the observer's sky which is known as the black hole shadow \cite{Cunha:2018acu,Vries_1999,Gralla:2019xty,Abdujabbarov:2015xqa,Abdujabbarov:2016hnw}. The outline of the shadow bears imprints of strong gravitational lensing of nearby radiation such that the shape and size of the shadow can unravel useful information regarding the nature of the background metric \cite{Gralla:2019xty,Bambi:2019tjh,Hioki:2009na,Vagnozzi:2019apd,Banerjee:2019nnj,Banerjee:2022iok}. 

In order to compute the outline of the shadow we need to adopt the Hamilton-Jacobi approach. 
This approach allows us to easily identify the constants of motion arising from the symmetries of the metric. The derivation of the shadow outline has been discussed in detail in \cite{Banerjee:2022iok}, therefore here we present only the main results. 
We start with the Lagrangian for particle motion $\mathcal{
L}(x^{\mu},\dot{x}^{\mu})=\dfrac{1}{2}g_{\mu\nu}\dot{x}^{\mu}\dot{x}^{\nu}$, from where it can be shown that the Hamiltonian satisfies the Hamilton-Jacobi equation,

\begin{flalign}
\mathcal{H} \left(x^{\mu},\dfrac{\partial S}{\partial x^{\mu}}\right) +\dfrac{\partial S}{\partial \lambda} =0 ;
\quad \dfrac{\partial S}{\partial x^{\mu}}=p_{\mu}
\label{3-1}
\end{flalign}
where the four momentum $p_{\mu}$ is connected with the Hamiltonian as $\mathcal{H} = \dfrac{1}{2} g^{\mu\nu} p_{\mu}p_{\nu} = \dfrac{k}{2}$, $k$ being the rest mass of the test particle. In \ref{3-1} $S$ represents the action and $\lambda$ the curve parameter.
Further, due to the symmetry of given space-time (in our case a stationary and axisymmetric space-time) we note that the specific energy E and the angular momentum L are the constants of motion given by,
\begin{flalign}
p_{t}& = -E = \rm constant\\
p_{\phi} &= L = \rm constant
\label{3-2}
\end{flalign}
Using \ref{3-1} in the on-shell condition $p^\mu p_\mu=0$ we note that the action is completely separable in all the four coordinates 
\begin{flalign}\label{HJ_SP4}
S= -E t +L \phi +S^{(r)}(r) + S^{\theta} (\theta)
\end{flalign}
This is attributed to the presence of Killing vectors ($\partial_t$, $\partial_\phi$) and a Killing tensor corresponding to the spacetime in \ref{Eq8}. 
The two unknown functions $S^r$ and $S^\theta$ can be obtained by solving the following differential equations,
\begin{flalign}
\left(\dfrac{dS^{r}}{dr} \right)^{2} = \dfrac{R(r)}{\Delta^{2}}\\
\left(\dfrac{dS^{\theta}}{d\theta}\right)^{2} = \Theta (\theta)
\label{3-3}
\end{flalign}
where 
\begin{flalign}
R(r) =& \Delta \left[ -C -(L-aE)^{2}\right]+ \left\{ \left(r^{2} +a^{2} \right)E -aL\right\}^{2} \\
\Theta(\theta) =& C +\cos^{2}\theta \left[ E^{2} a^{2} -\dfrac{L^{2}}{\sin^{2}\theta}\right]
\label{3-4} 
\end{flalign}
In \ref{3-4} $C$ refers to the third constant of motion and is known as the Carter constant.
Therefore, the solution of the Hamilton-Jacobi equation has the form
\begin{flalign}
S=  -E t + L\phi + \int \dfrac{\sqrt{R(r)}}{\Delta} dr + \int \sqrt{\Theta(\theta)} d\theta
\label{3-5}
\end{flalign}
from where one can show that the equations of motion for $r$ and $\theta$ are given by
\begin{subequations}
\begin{flalign}
\dot{r} &= \dfrac{\sqrt{R(r)}}{\Sigma} \label{SP4_D1} \\ 
\dot{\theta} &=\dfrac{\sqrt{\Theta}}{\Sigma}\label{SP4_D2}
\end{flalign}
\end{subequations}
From the constants of motion we can construct the two impact parameters
\begin{flalign}\label{SP4_IP}
\chi=\dfrac{C}{E^{2}} \quad \textit{\rm and} \quad \eta=\dfrac{L}{E}
\end{flalign}
which represents the distance from the equatorial plane and the axis ratio respectively. 
In the next section we briefly discuss the nature of motion of photons in the radial and vertical direction.

\subsection{Motion of photons in the vertical and radial direction}
At first we will deal with the angular part and hence will solve \ref{SP4_D2}. For solving the same we introduce the new variable $u=\cos\theta$. In terms of this new variable the differential equation for angular part becomes
\begin{flalign}\label{SP4_Ang}
\left(\dfrac{\Sigma}{E}\right)^{2}\dot{u}^{2}=\chi -u^{2}(\chi+\eta^{2}-a^{2})-a^{2}u^{4} \equiv F(u(\theta))
\end{flalign}
Above equation (\ref{SP4_Ang}) indicates that for physically realisable solution, $F(u)$ cannot be negative. However, when $u=1$ i.e $\theta=0^{\circ}$ the function $F(u)$ becomes negative which suggests that the photon cannot access the full angular space. If one wish to find the maximum angular space available i.e $\theta_{max}$ one need to set $F(u_{0})=0$. One can solve this equation and get
\begin{flalign}
u_{0}^{2}=\dfrac{-(\chi+\eta^{2}-a^{2})\pm \sqrt{(\chi+\eta^{2}-a^{2})^{2}+4a^{2}\chi}}{2a^{2}}
\label{3-6}
\end{flalign}
Depending on the signature of $\chi$ the photons can access certain regions of angular space. For example, when $\chi>0$, only the positive root of \ref{3-6} is relevant. In this case the physically allowed range of $u$ lies between $\pm |u_0|$ such that the orbits cross the equatorial plane. When $\chi=0$, we can have two roots,
\begin{align}
u_1^2=0~~~~~~~~~~~~~~~~~~~~~~\rm{and} ~~~~~~~~~~~~~~~~~~~~~~~~u_2^2=1-\frac{\eta^2}{a^2}
\label{3-7}
\end{align}
and when $\eta^2>a^2$ only the first root is allowed. 
When $\chi$ is negative, \ref{3-6} can be written as,
\begin{align}
u_0^2=\frac{|\chi|-\eta^2+a^2\pm \sqrt{(-|\chi|+\eta^2-a^2)^2-4a^2|\chi|})}{2a^2}
\label{3-8}
\end{align}
where $\chi=-|\chi|$. It is evident from \ref{3-8} that $u_0$ has a physically realizable solution when the following condition is satisfied,
\begin{align}
a^2 + |\chi|-\eta^2 >0
\label{3-9}
\end{align}

Differential equation associated to the radial part is given by
\begin{flalign}
 \left(\dfrac{\Sigma}{E}\right)^{2} \dot{r}^{2}  =\Delta\left[-\chi +\dfrac{k}{E^{2}}r^{2} -\left(\dfrac{L}{E}-a\right)^{2}\right]+\left[r^{2}+a^{2}-a\eta\right]^{2} \equiv V(r)
\end{flalign}
%%%%%%%%%%%%%
Spherical photon orbits are realised with the conditions $V(r)=0=V'(r)$. These conditions further determine two parametric equations
\begin{subequations}
\begin{equation}
 \Delta \left(\chi + \eta^{2}+a^{2}-2\eta a \right)=\left[r^{2}+a^{2}-a\eta\right]^{2}
 \end{equation}
 \begin{equation}
 {\chi +\eta^{2}+a^{2}-2\eta a} = \dfrac{2}{ \left[1-\dfrac{m(r)}{r}-m'(r)\right]}\left[r^{2}+a^{2}-a\eta\right]
 \end{equation}\label{SP4_cond}
\end{subequations}
Solving \ref{SP4_cond} we immediately land with two sets of solutions for the impact parameters as given below
\begin{enumerate}
\item 
\begin{subequations}
\begin{equation}
\chi = -\dfrac{r^{4}}{a^{2}}
\end{equation}
\begin{equation}
\eta = \frac{a^2 + r^{2}}{a }
\end{equation}\label{SP4_cond_1}
\end{subequations}

\item  
\begin{subequations}
\begin{flalign}
\chi&=-\frac{r^3 \left(4 a^2 r m'(r)-4 a^2 m(r)+r^3 m'(r)^2+2 r^3 m'(r)-6 r^2 m(r) m'(r)-6 r^2 m(r)+9 r m(r)^2+r^3\right)}{a^2 \left(r m'(r)+m(r)-r\right)^2}\\
\eta&= \frac{a^2 r m'(r)+a^2 m(r)+a^2 r+r^3 m'(r)-3 r^2 m(r)+r^3}{a \left(r m'(r)+m(r)-r\right)}
\end{flalign}\label{SP4_Cond_2}
\end{subequations}
 \end{enumerate}
First solution is physically not allowed because it does not satisfy \ref{3-9}. For the second solution $\chi$  may take either sign depending upon the value of $r$ (where $r$ is associated with the radius of the spherical photon orbits). But in general
\begin{flalign}
&a^{2}-\chi-\eta^{2} = -\frac{2 r \left(\left\{r m'(r)+m(r)\right\} \left\{2 a^2+r^2 m'(r)-3 r m(r)\right\}+r^3\right)}{\left\{r m'(r)-r+m(r)\right\}^2}
\end{flalign}
and $\chi=0$ refers to case of motion in the equatorial plane.

\subsection{Shape of the black hole shadow}
\label{S3-2}
\begin{figure}[h!]
\centering
\includegraphics[scale=.3]{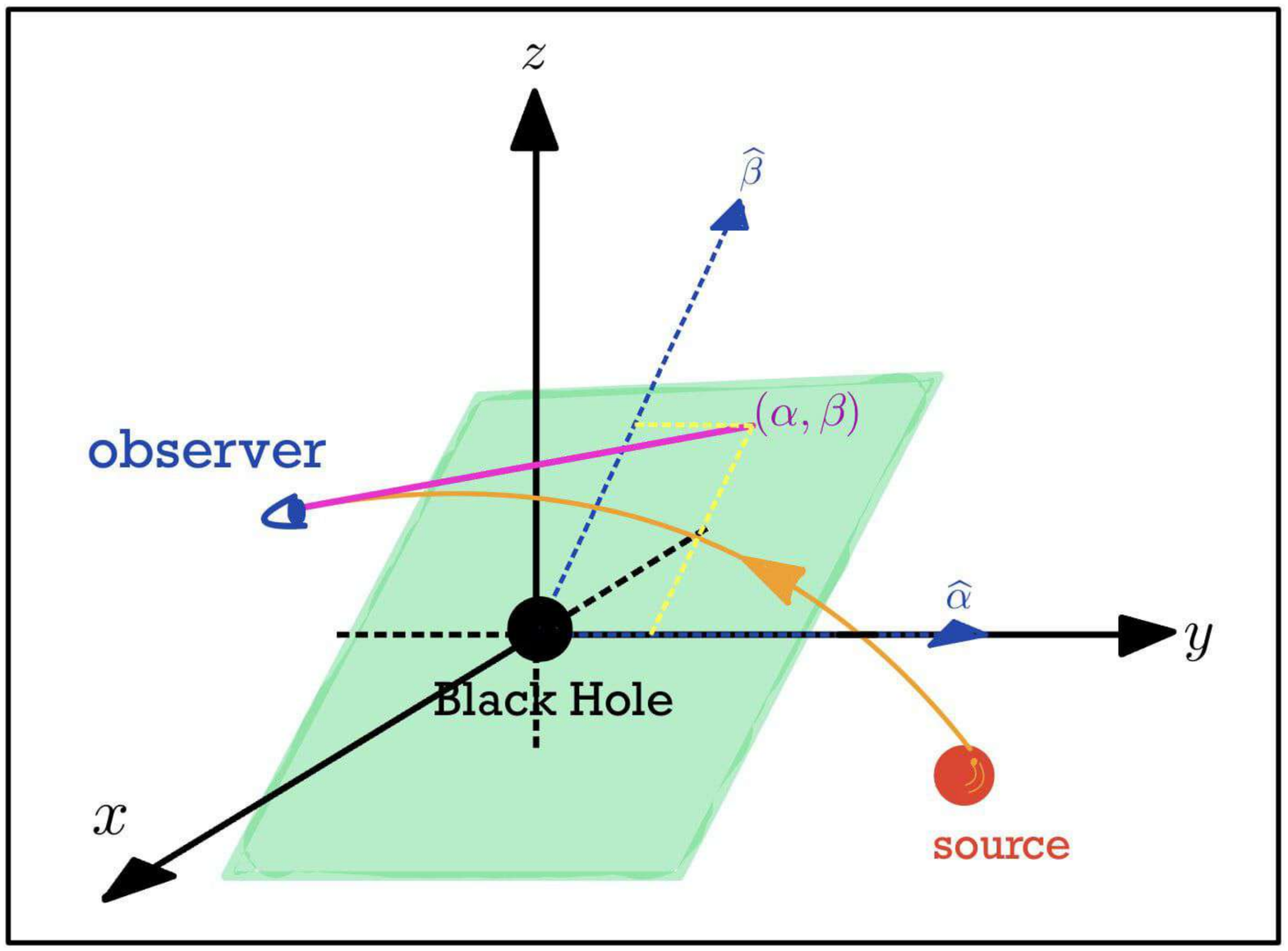}
\caption{In this figure geometry of gravitational lens has been shown. Any observer far away from black hole can imagine a reference coordinate system $(x,y,z)$ with the black hole at its origin. At infinity the Boyer-Lindquist coordinates coincides with the reference coordinate system and as seen from infinity the black hole rotates around $z$-axis. The line connecting the origin and the observer is normal to the $\alpha-\beta$ plane.}
\label{Fig1}
\end{figure}

As the space-time that we have considered in our present scenario is asymptotically flat, a distant observer can imagine a reference Euclidean coordinate system $(x,y,z)$ with black hole at its origin. One should note that the Boyer-Lindquist coordinate becomes identical to the reference Euclidean coordinate system for large values of $r$. Further we have also assumed that as seen from infinity the axis of rotation of black hole is along $z-$axis. Without loss of generality, let us assume the coordinate of the distant observer is $(r_{0},\theta_{0},0)$. In a similar way $(r_{s},\theta_{s},\phi_{s})$ is assumed to be the position of source \cite{Vazquez:2003zm}.

Considering the observer's frame, an incoming light ray can be described by a parametric curve ${x(r), y(r), z(r)}$ such that $r^{2}=x^{2}+y^{2}+z^{2}$. One can note that this is just the usual radial coordinate in BL coordinate system if $r$ is considerably large. At the location of the observer, the tangent vector $\left(\vec{v}\big|_{(r_{0},\theta_{0},0)}\right)$ to the parametric curve can be provided as follows
\begin{flalign}
\vec{v}\big|_{{(r_{0},\theta_{0},0)}}=\dfrac{dx}{dr}\bigg|_{{(r_{0},\theta_{0},0)}}\hat{x}+\dfrac{dy}{dr}\bigg|_{{(r_{0},\theta_{0},0)}}\hat{y}+\dfrac{dz}{dr}\bigg|_{{(r_{0},\theta_{0},0)}}\hat{z} 
\end{flalign}
 This tangent vector represents a straight line which intersects the $\alpha-\beta$ plane at $(\alpha_{i},\beta_{i})$. One should note that the line connecting the origin and the observer is normal to this $\alpha-\beta$ plane. This $\alpha-\beta$ plane is known as observer's (view) plane. This is schematically shown in \ref{Fig1}.

Using usual coordinate transformation rule and for the observer considering $\phi=0$, we have the position of observer as follows
\begin{flalign}
x_{o}&=r_{o}\sin\theta_{o} \nonumber\\
y_{o}&=0\nonumber\\
z_{o}&=r_{o}\cos\theta_{o} \nonumber
\end{flalign}

Using coordinate transformation one can easily find the components of the  tangent vector at the position of the observer. These are given by
%\item
\begin{subequations}
 \begin{flalign}
\dfrac{dx}{dr}\bigg|_{\vec{r}_{0}}&=\sin\theta_{0}+r_{0}\cos\theta_{0}\dfrac{d\theta}{dr}\bigg|_{\vec{r}_{0}} \\
\dfrac{dy}{dr}\bigg|_{\vec{r}_{0}}&=r_{0}\sin\theta_{0}\dfrac{d\phi}{dr}\bigg|_{\vec{r}_{0}}\\
\dfrac{dz}{dr}\bigg|_{\vec{r}_{0}}&=\cos\theta_{0}-r _{0}\sin\theta_{0}\dfrac{d\theta}{dr}\bigg|_{\vec{r}_{0}}
\end{flalign}
\end{subequations}

%\item 

\begin{figure}[h!]%
    \centering
   \subfloat[Variation of BH shadow with metric parameter $g$. The inclination angle and spin parameter are taken to be respectively $\theta=45^{\circ}$ and $a=0.2$. ]{{\includegraphics[width=7.5cm]{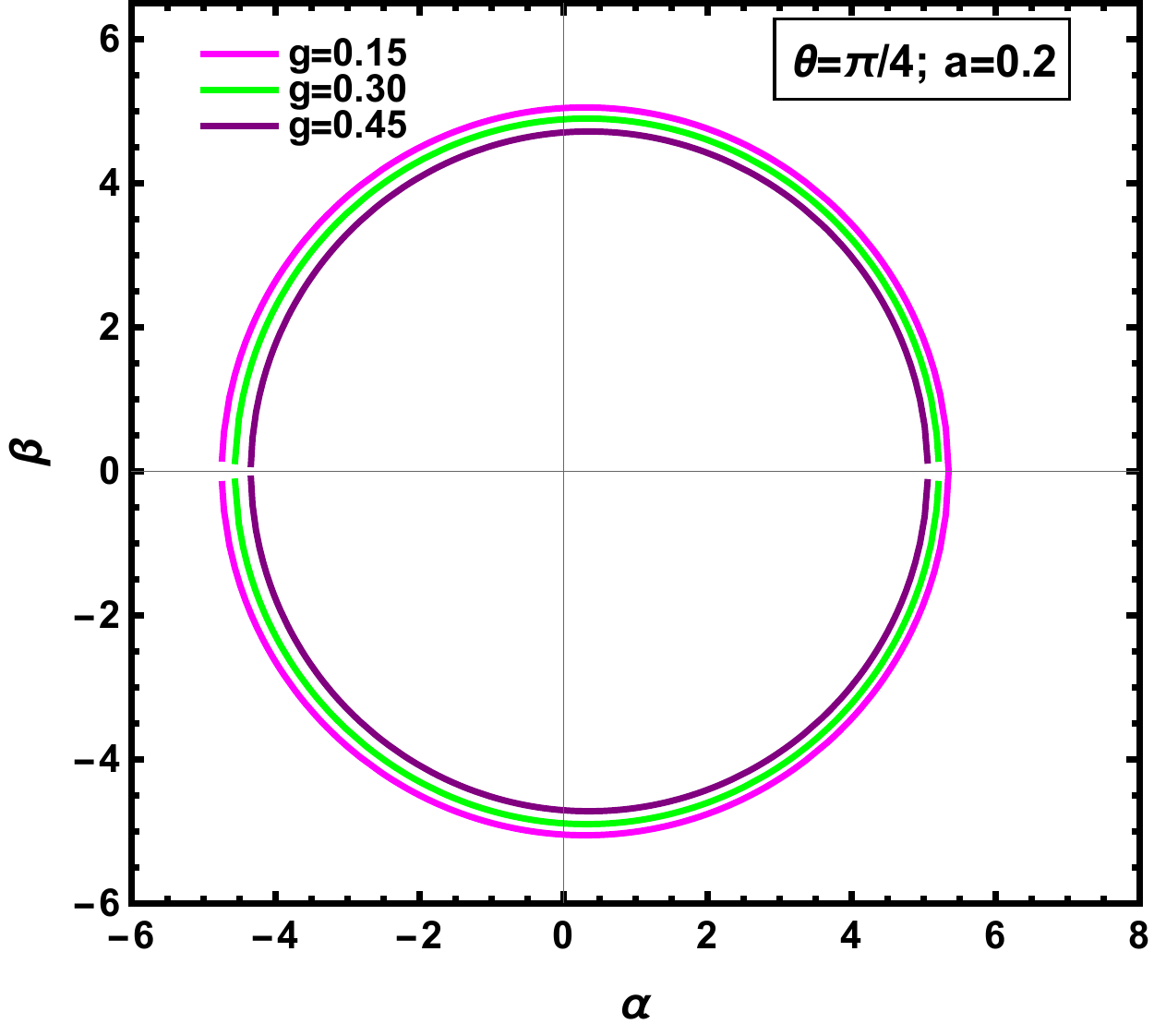} } \label{SP4_Fig_Vg_a}}
   \qquad
    \subfloat[Variation of shadow with metric parameter $g$.The inclination angle and spin parameter are taken to be respectively $\theta=60^{\circ}$ and $a=0.5$. ]{{\includegraphics[width=7.5cm]{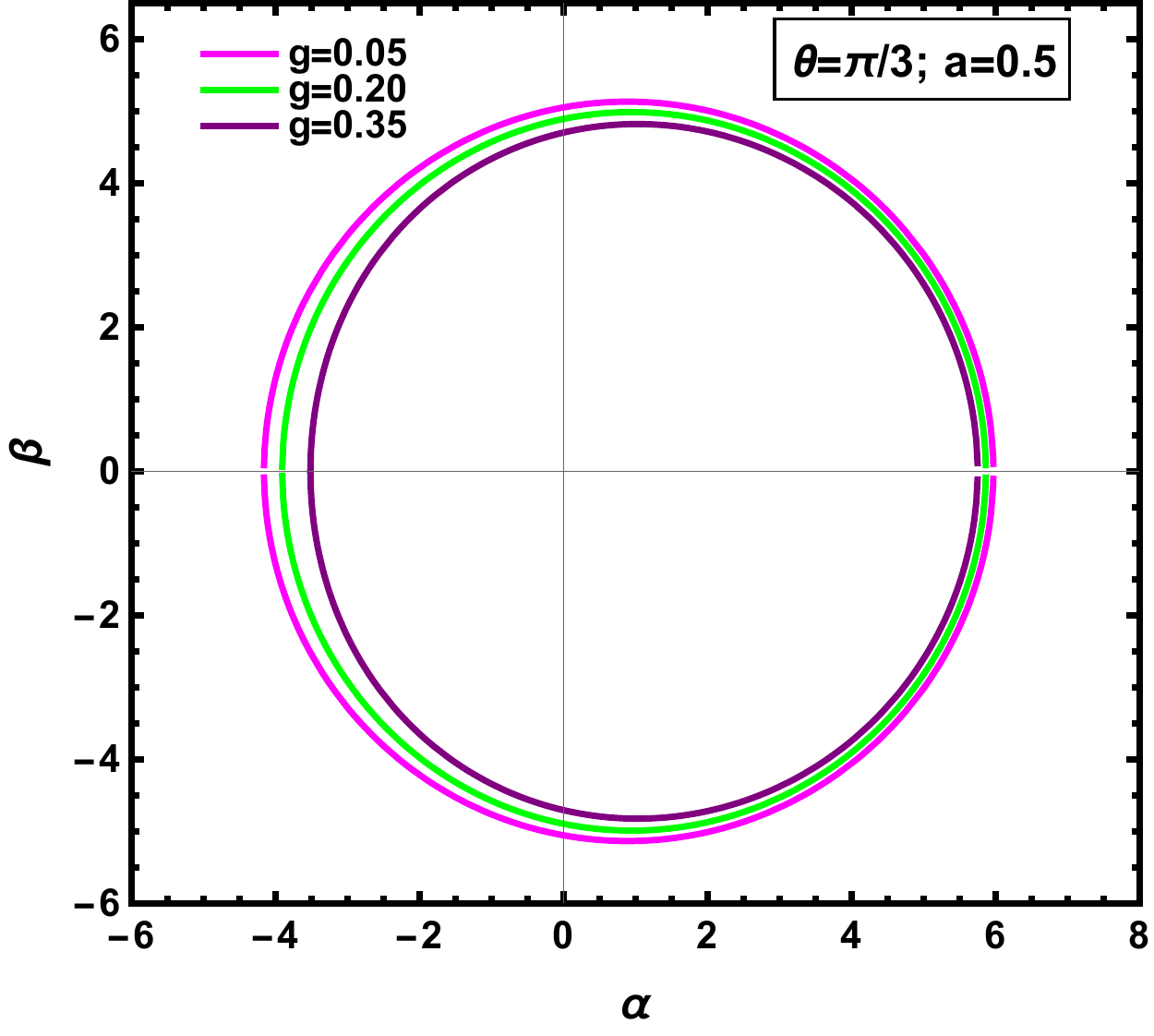} } \label{SP4_Fig_Vg_b}}
    \qquad
    \hspace{-1.5cm}
    \subfloat[Variation of shadow with spin-parameter $a$. Here the inclination angle is $\theta=60^{\circ}$ and $g=0.1$.]{{\includegraphics[width=7.5cm]{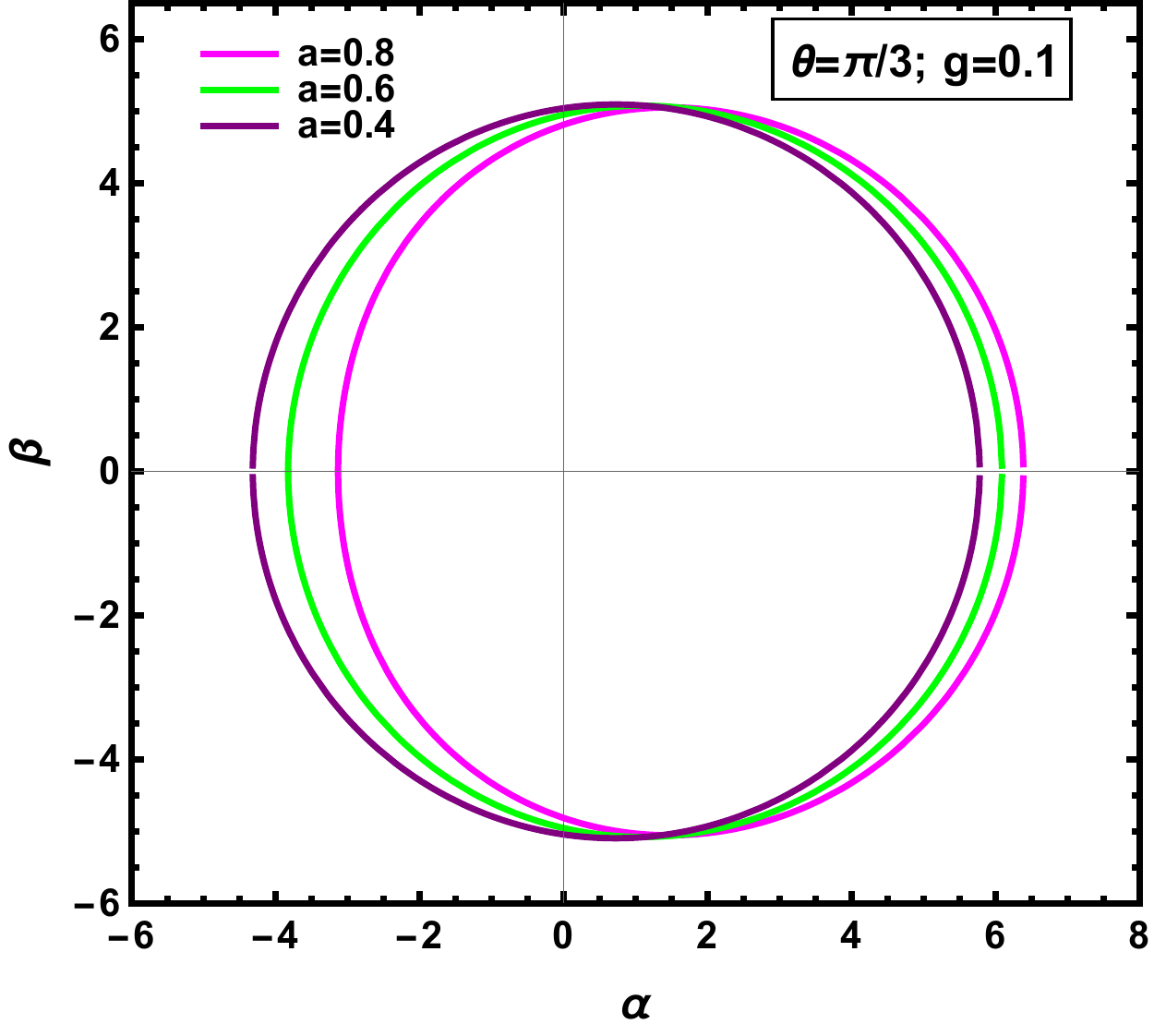} } \label{SP4_Fig_Vg_c}}
\qquad    
    \subfloat[Variation of shadow structure with different inclination angle $\theta$.Here the spin is $a=0.8$ and $g=0.1$.]{{\includegraphics[width=7.5cm]{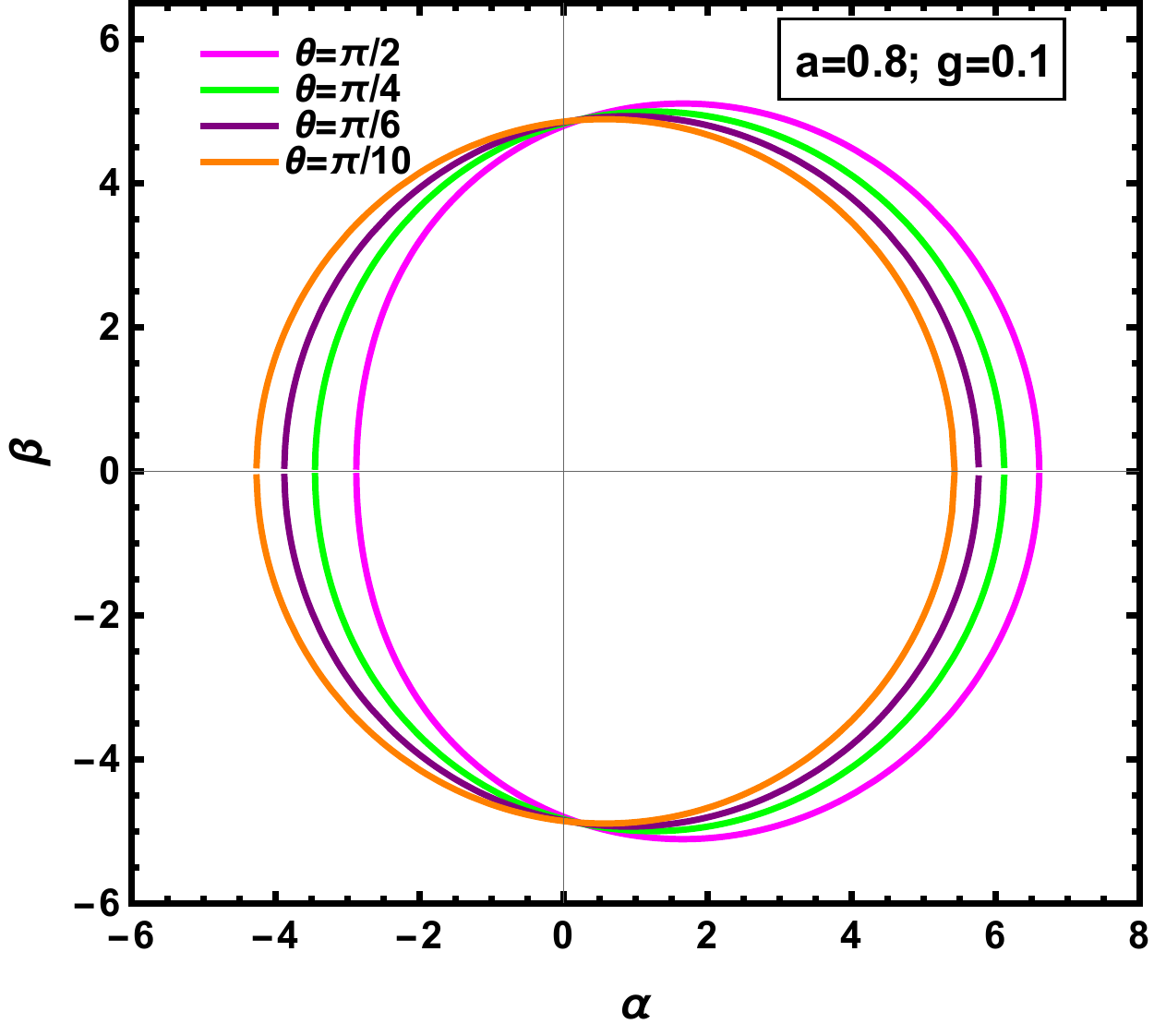} } \label{SP4_Fig_Vg_d}}
   \caption{Variation of shape and size of black hole shadow with magnetic monopole charge parameter $g$, rotation parameter $a$ and inclination angle $\theta$.}
    \label{SP4_Fig_S1_Vg}
\end{figure}

This vector describes a straight line which intersect the  $\alpha-\beta$ plane.The point $(\alpha_{i},\beta_{i})$ in the plane $(\alpha,\beta)$ in $(x,y,z)$ system is given by
\begin{flalign}
x&=-\beta_{i}\cos\theta_{0}\nonumber\\
y&=\alpha_{i}\nonumber\\
z&=\beta_{i}\sin\theta_{0}\nonumber
\end{flalign}
Now
\begin{flalign}
\vec{r}_{0}-r_{0}\vec{v}\bigg|_{\vec{r}_{0}}=\left( r_{0}\sin\theta_{0}-r_{0}\dfrac{dx}{dr}\bigg|_{\vec{r}_{0}}, -r_{0}\dfrac{dy}{dr}\bigg|_{\vec{r}_{0}},r_{0}\cos\theta_{0}-r_{0}\dfrac{dz}{dr}\bigg|_{\vec{r}_{0}} \right)
\end{flalign}
Hence we have
\begin{flalign}
-\beta_{i}\cos\theta_{0}&=r_{0}\sin\theta_{0}-r_{0}\dfrac{dx}{dr}\bigg|_{\vec{r}_{0}}=r_{0}\sin\theta_{0}-r_{0}\left\{\sin\theta_{0}+r_{0}\cos\theta_{0}\dfrac{d\theta}{dr}\bigg|_{\vec{r}_{0}} \right\}\nonumber\\
\implies \beta_{i}&=r_{0}^{2}\dfrac{d\theta}{dr}\bigg|_{\vec{r}_{0}}
\end{flalign}
Similarly we have
\begin{flalign}
\alpha_{i}=-r_{0}\dfrac{dy}{dr}\bigg|_{\vec{r}_{0}}=-r_{0}^{2}\sin\theta_{0}\dfrac{d\phi}{dr}\bigg|_{\vec{r}_{0}}
\end{flalign}

For the sake of visualization of black hole shadow, let us consider a distant observer at position $(r_{0},\theta_{0})$. If $\{\alpha_{i},\beta_{i}\}$ are the  celestial coordinates then the stereographic projection of the shadow from celestial body to the image plane can be defined as follows \cite{Vazquez:2003zm}
\begin{subequations}
\begin{flalign}
\alpha=&\lim_{r_{0}\to\infty}\left(-r_{0}^{2}\sin\theta_{0}\dfrac{d\phi}{dr} \right) \\
\beta=&\lim_{r_{0}\to\infty}\left(r_{0}^{2}\dfrac{d\theta}{dr} \right)
\end{flalign}
\end{subequations}

In terms of impact parameters, as defined in \ref{SP4_IP}  these coordinates are given by 

\begin{flalign}
\alpha_{i}&= - \eta \csc\theta_{o}\\
\beta_{i} & = \sqrt{\chi + a^{2}\cos^{2}\theta_{0}-\eta^{2}\cot^{2}\theta_{0}}
\end{flalign}
 
In \ref{SP4_Fig_S1_Vg} we have plotted the variation of size and shape of black hole shadow with change in the charge parameter $g$, the rotation parameter $a$ and the inclination angle $\theta$. From \ref{SP4_Fig_Vg_a}
and \ref{SP4_Fig_Vg_b} it can be noted that for a fixed value of inclination angle the size of the black hole shadow gradually decreases for an increase in non-linear electrodynamics charge parameter $g$. From  \ref{SP4_Fig_Vg_c} and \ref{SP4_Fig_Vg_d} one can notice that the shape of the black hole shadow becomes more dented and deviates from circular shape for both an increase in spin parameter $a$ and inclination angle $\theta$.

\section{Comparison with the observed shadow of M87* and Sgr A*}
\label{SP4_S4}
In this section we compare the theoretically derived shadow with the observed image of M87* and Sgr A*. In order to accomplish this we need to define a few observables associated with black hole shadow.
The boundary of the shadow is given by the curve $\beta(\alpha)$. As the shadow is not circular in shape, one need to define one major axis and one minor axis associated with two diameters as shown in \ref{Fig_IMG}.

 The major axis is introduced by $\Delta\beta=\beta_{t}-\beta_{b}$, where $\beta_{t}$ and $\beta_{b}$ are  the top-most and the bottom-most points of the shadow respectively in $\alpha-\beta$ plane. In a similar manner the minor axis is given  by $\Delta \alpha=\alpha_{r}-\alpha_{l}$, where $\alpha_{r}$ and $\alpha_{l}$ are  the right-most and left-most points  of the shadow respectively. 
 
 One can note from \ref{SP4_Fig_S1_Vg} that these two axes i.e $\Delta\beta$ and $\Delta\alpha$ can also be expressed as a function of metric parameters ($g$ and $a$) and the inclination angle.  The shadow is symmetric in nature about the minor axis. We next discuss some of the observables which are defined to gain a theoretical understanding regarding the effect of modification in the gravity theory:
\begin{enumerate}[label=(\roman*)]
\item Angular diameter which is given by
\begin{flalign}\label{SP4_Phi}
\Phi=\dfrac{GM}{c^{2}D}\Delta \beta
\end{flalign}
It is important to note that the theoretical angular diameter depends on the mass $M$ of the black hole and the distance $D$ of the black hole from the observer. One therefore requires independent measurement of the mass and distance to evaluate the theoretical angular diameter which is then compared with the observations.
\item Axis ratio that is given by
\begin{flalign}\label{SP4_A}
\Delta A=\dfrac{\Delta\beta}{\Delta\alpha}
\end{flalign}

\begin{figure}[t!]
\centering
\includegraphics[scale=.7]{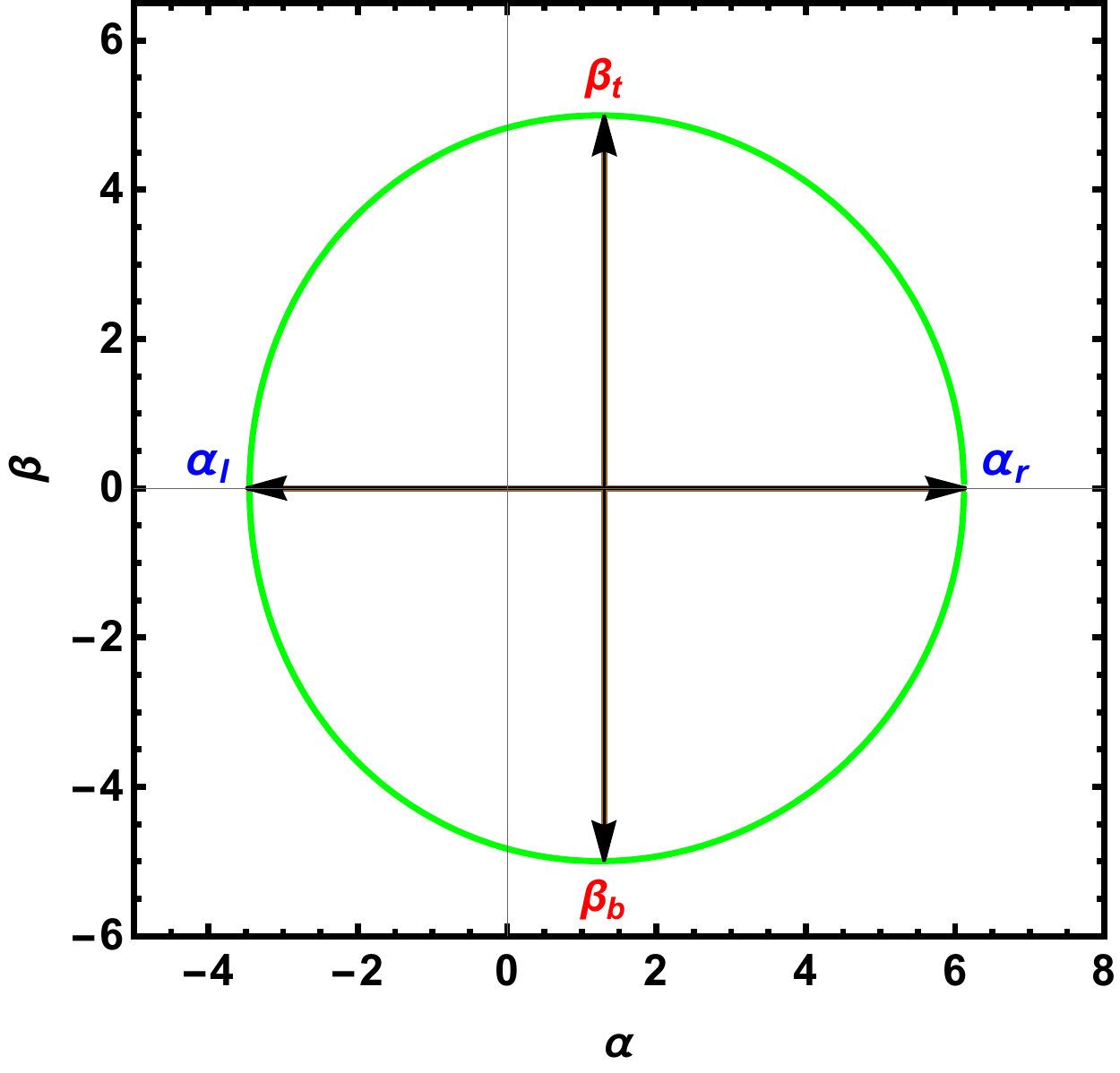}
\caption{Major axis and minor axis of shadow structure}
\label{Fig_IMG}
\end{figure}
\item Deviation from circularity: Another important observable associated with the black hole shadow is the deviation from circularity. Unlike shadow of spherically symmetric metric, the shadow structure for Kerr-like solution has an appearance resembling the dented circular shape and that is why one needs to define the deviation from the circular shape of the shadow. The deviation from circularity is defined as
\begin{flalign}
\Delta C= \dfrac{1}{R_{\rm avg}}\sqrt{\dfrac{1}{2\pi}\int_{0}^{2\pi}d\phi\left\{ \ell(\phi)-R_{\rm avg}\right\}}
\end{flalign}
where $\ell(\phi)=\sqrt{\left\{\alpha(\phi)-\alpha_{c} \right\}^{2}+\beta^{2}(\phi)}$ corresponds to the distance between the geometrical centre and any point $(\alpha,\beta)$ on the shadow curve with azimuthal angle $\phi$. The geometric centre for shadow is given by $\alpha_{c}=\int \alpha d\mathcal{A}/ \int d\mathcal{A}$ and $\beta_{c}=0$ (due to reflection symmetry around minor axis), where $d\mathcal{A}$ is the area element while the average radius
$R_{\rm avg}$ assumes the form,
\begin{flalign}
R_{\rm avg}=\sqrt{\dfrac{1}{2\pi}\int_{0}^{2\pi}d\phi\, \ell^{2}(\phi)}
\end{flalign}
\end{enumerate}
 
\subsection{Constrains on the magnetic monopole charge from the image of M87* }
In this section, we compare the theoretically computed shadow outline with the observed image of M87*. Observational results of the Event Horizon Telescope (EHT) collaboration \cite{Akiyama:2019cqa,Akiyama:2019fyp,Akiyama:2019eap} has imposed bounds on the angular diameter, the deviation from circularity and the axis ratio. These bounds are:
\begin{enumerate}
\item Angular diameter: $\Phi=(42\pm3)\mu as$
\item Deviation from circularity: $\Delta C \lesssim 10\%$
\item Axis ratio: $\Delta A\lesssim 4/3$
\end{enumerate}

\begin{figure}[h!]%
    \centering
     \subfloat[Contours represent the angular diameter of the shadow of M87* calculated with mass $M\simeq 3.5\times 10^{9}M_{\odot}$ and distance $D\simeq 16.8 \rm Mpc$]{{\includegraphics[width=7.5cm]{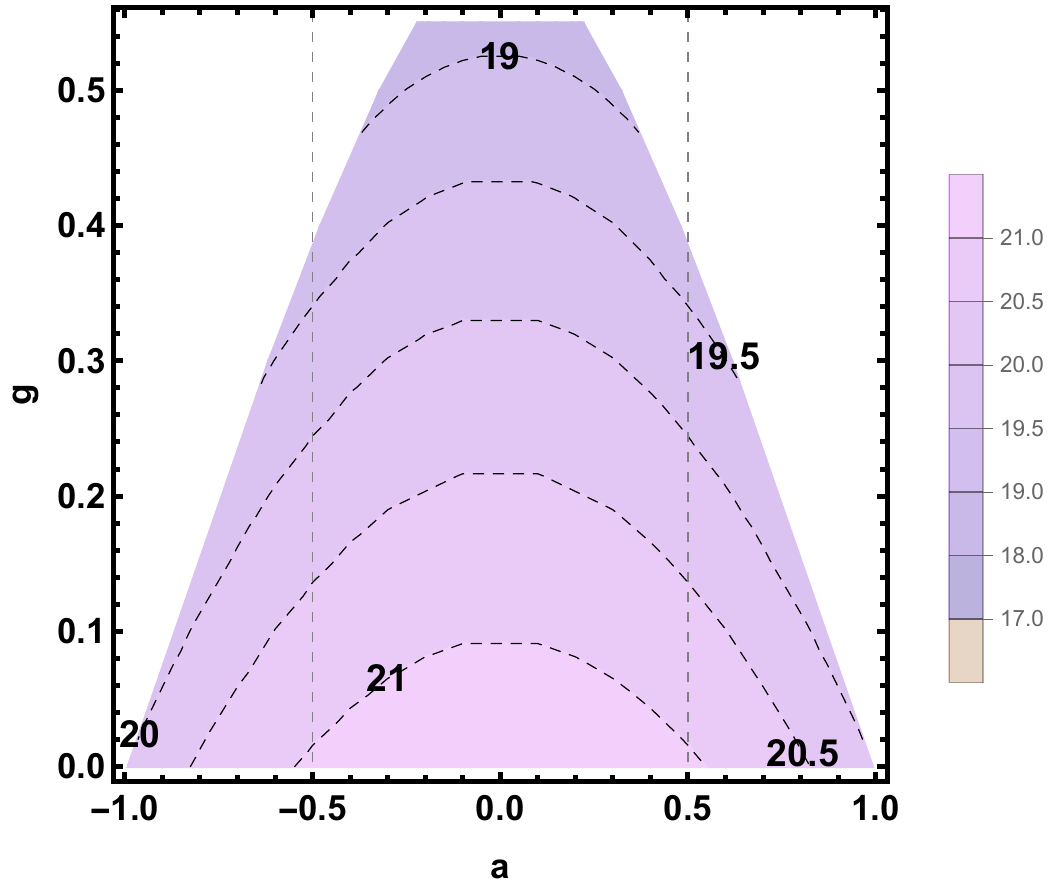} } \label{SP4_Fig_M1a}}
    \qquad
    \subfloat[Contours represent the angular diameter of the shadow of M87* calculated with mass $M\simeq 6.2\times 10^{9}M_{\odot}$ and distance $D\simeq 16.8 \rm Mpc$]{{\includegraphics[width=7.5cm]{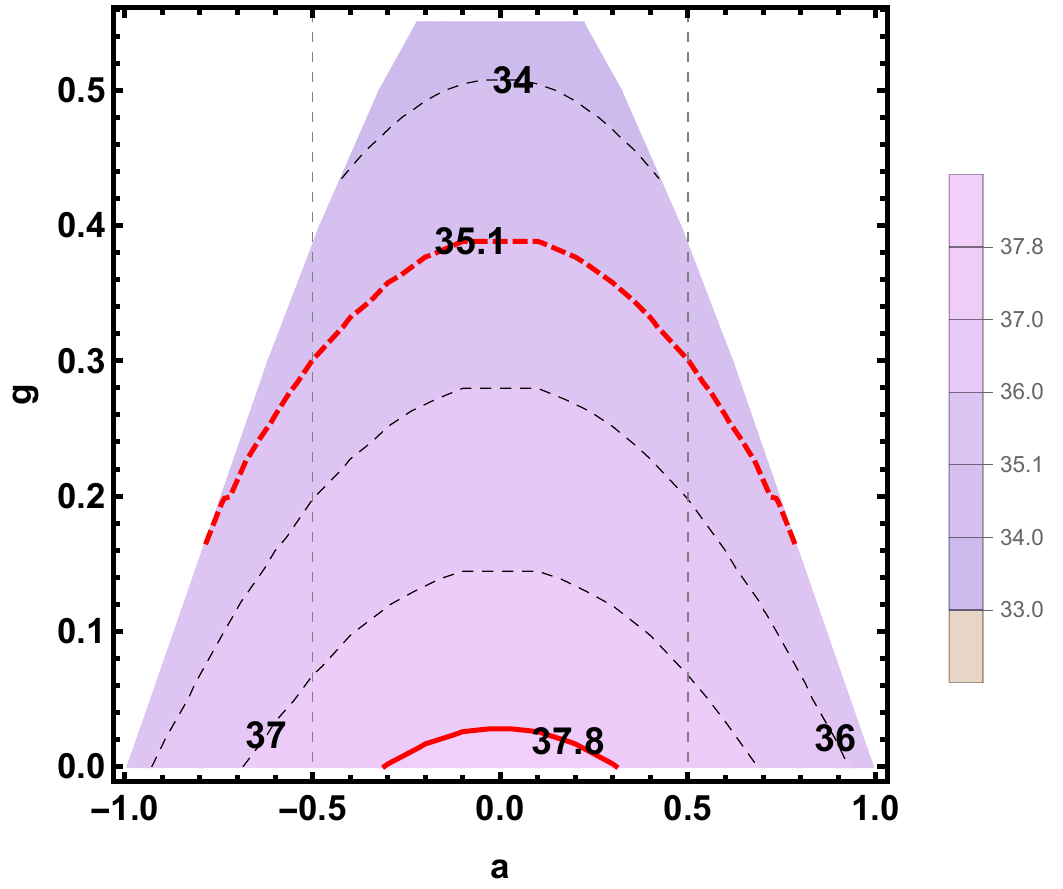} }
    \label{SP4_Fig_M1b}}
    \caption{Variation of the angular diameter of M87* in $(a,g)$ plane has been depicted. }
    \label{SP4_Fig_M1}%
\end{figure}

\begin{figure}
\centering
{\includegraphics[width=8cm]{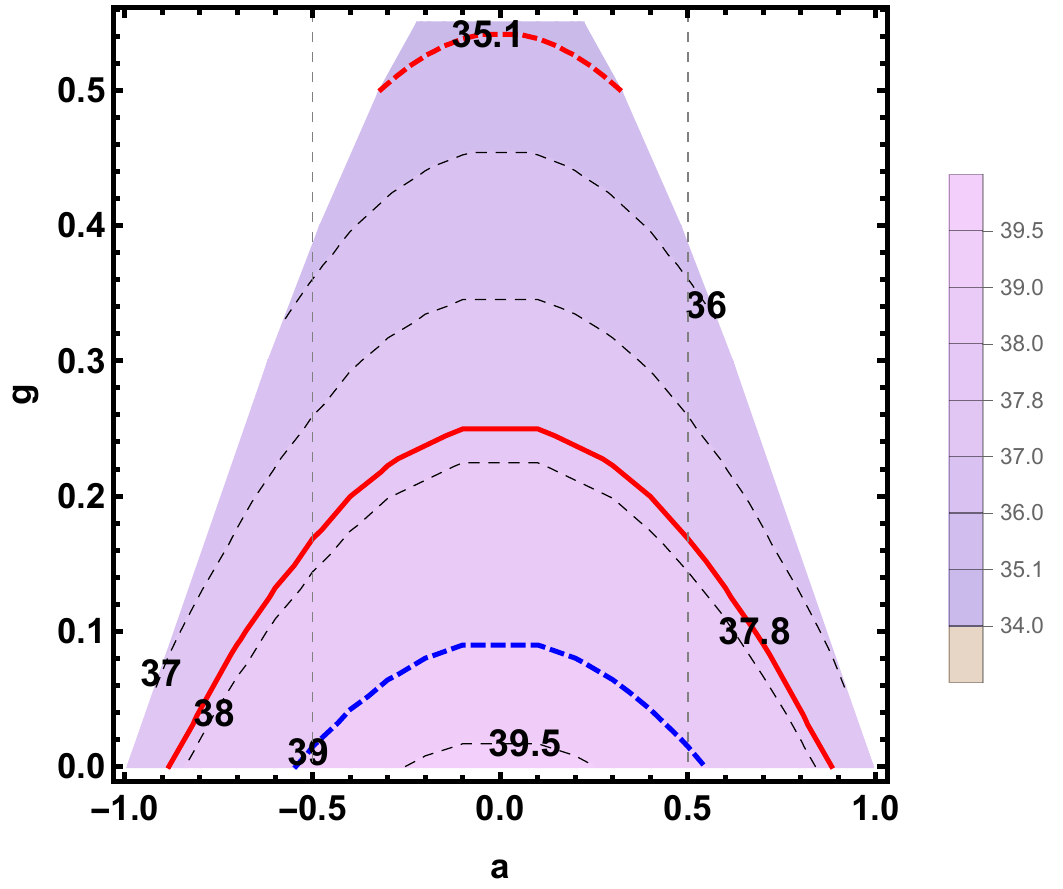}}
\caption{The variation of angular diameter of shadow for M87* with $a$ and $g$ for mass $M\simeq 6.5\times10^{9} M_{\odot}$ and distance $D=16.8\rm Mpc$}
\label{SP4_Fig_M2}
\end{figure}

Analysing the motion of stars and gas clouds moving near the M87* super massive black hole, the mass of the source has been measured. Study of gas dynamics points out the mass of M87* to be $M \simeq 3.5_{-0.3}^{+0.9}\times 10^{9} M_{\odot}$ \cite{Walsh:2013uua}, whereas the stellar dynamics measurements suggest the mass to be $M \simeq 6.2_{-0.5}^{+1.1}\times 10^{9} M_{\odot}$ \cite{Gebhardt:2011yw}. The distance of M87* has been reported $D=(16.8\pm 0.8)\rm Mpc$ (from stellar population measurement) \cite{Blakeslee:2009tc,Bird:2010rd,Cantiello:2018ffy}.
Assuming the jet axis to coincide with the rotation axis, the inclination angle has been estimated to be $17^{0}$ \cite{Akiyama:2019cqa,Akiyama:2019fyp,Akiyama:2019eap}. Additionally, from the observed angular diameter of M87* EHT collaboration reports the mass of the black hole to be $M\approx 6.5\times 10^{9} M_{\odot}$.

\begin{figure}[h!]%
    \centering
     \subfloat[The above figure presents the variation of deviation from circularity $\Delta C$ with $g$ and $a$ assuming an inclination angle of $17^{\circ}$ corresponding to M87*]{{\includegraphics[width=7.5cm]{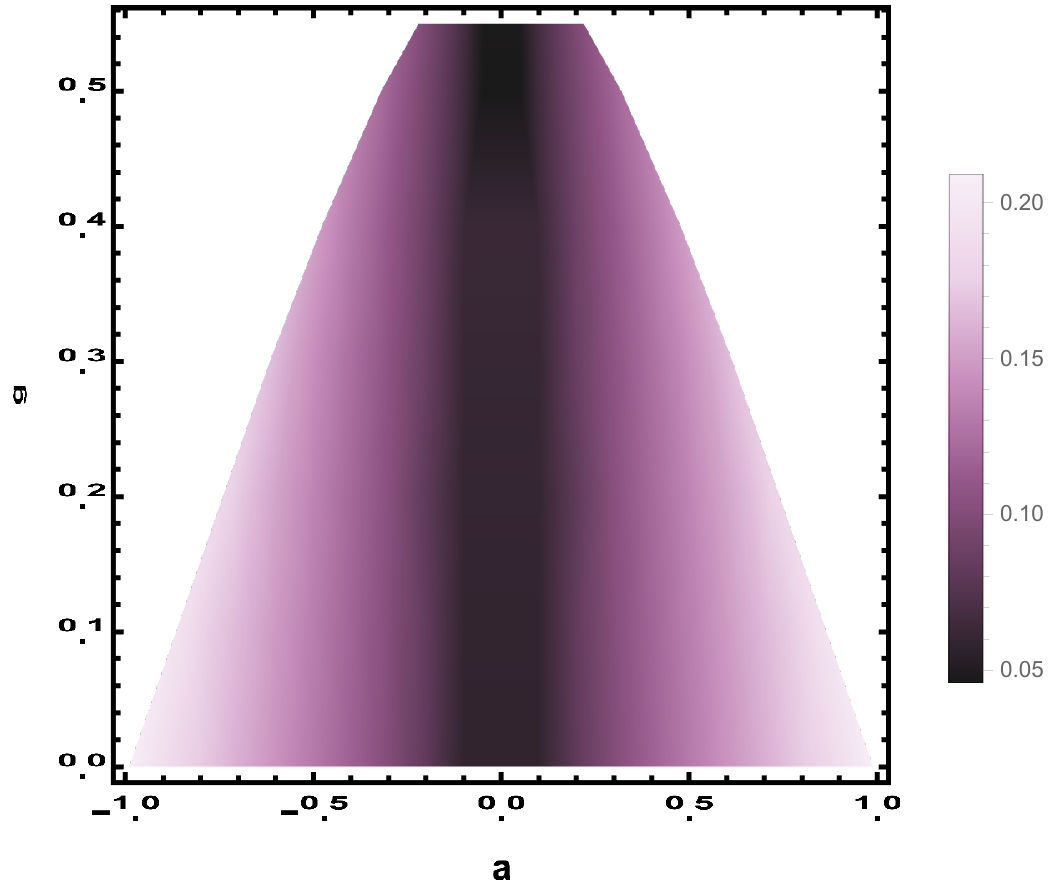} }
     \label{SP4_Fig_M3a}}
    \qquad
    \subfloat[The above figure presents the variation of axis ratio $\Delta A$ with $g$ and $a$ assuming an inclination angle of $17^{\circ}$ corresponding to M87*]{{\includegraphics[width=7.5cm]{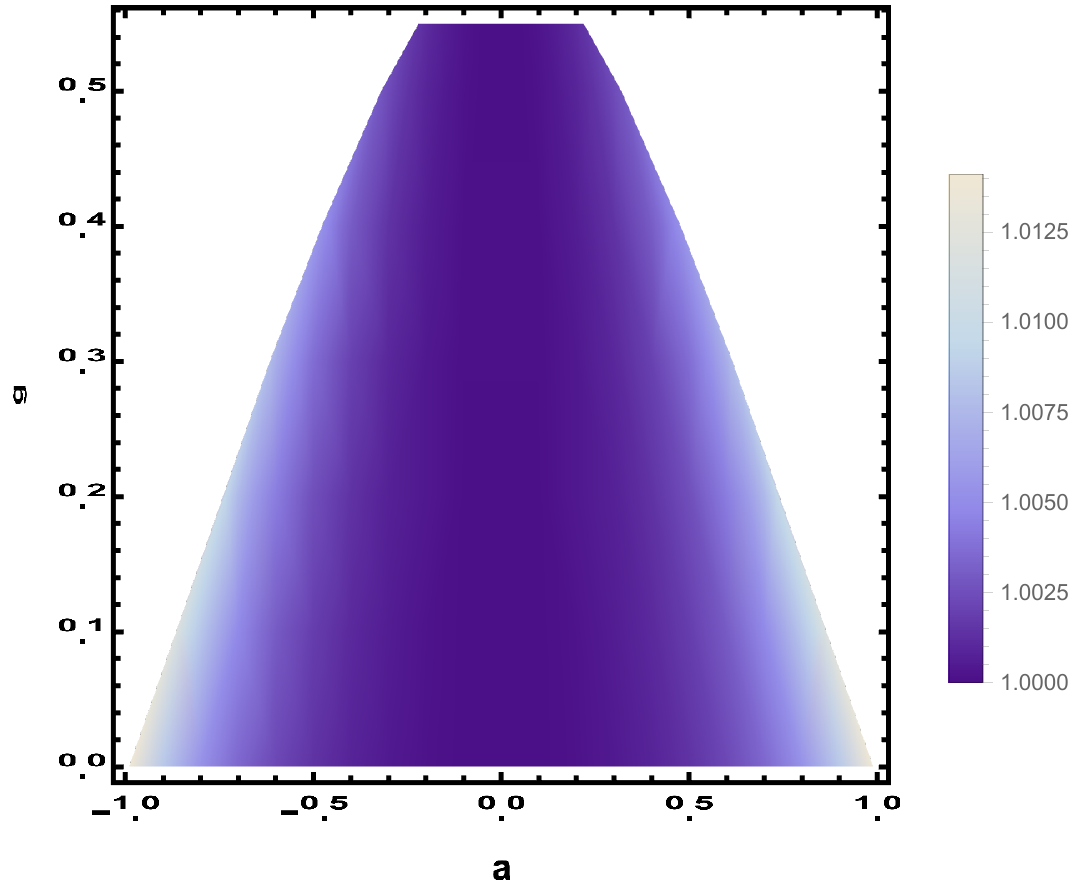} }
    \label{SP4_Fig_M3b}}
    \caption{In above two figures we have shown the variation of axis ratio $\Delta A$ and deviation from circularity $\Delta C$ as a function of metric parameter $g$ and spin parameter $a$. In these cases we have taken the inclination angle $17^{\circ}$  for M87* }
    \label{SP4_Fig_M3}%
\end{figure}

We recall at this point that independent measurements of mass and distance are required to obtain the theoretical angular diameter (see \ref{SP4_Phi}) whereas the dependence of $\Phi$ on the background metric is encoded in $\Delta\beta$ (where $\Delta\beta=\beta_{t}-\beta_{b}$ as shown in \ref{Fig_IMG}). In \ref{SP4_Fig_M1} and \ref{SP4_Fig_M2} the variation of the angular diameter has been shown with the change in the magnetic monopole charge parameter $g$ and the spin parameter $a$. In \ref{SP4_Fig_M1a} the angular diameter is calculated with mass $3.5\times 10^9 M_\odot$ while in \ref{SP4_Fig_M1b} the angular diameter is evaluated with mass $6.2\times 10^9 M_\odot$. In both cases the distance is taken to be $16.8$ Mpc. 
The observed angular diameter of M87* reported by the Event Horizon Telescope collaboration turns out to be $(42\pm 3)\mu as$. The EHT collaboration also states that the observed image can have a maximum offset of $10\%$ which implies that the shadow of M87* can be as small as 
$(37.8\pm2.7)\mu as$. From \ref{SP4_Fig_M1a} it is clear that with mass $3.5\times 10^9 M_\odot$ the observed angular diameter cannot be reproduced. When $M\simeq 6.2 \times 10^9 M_\odot$ is used, the maximum value of $g$ that can explain the observed angular diameter with $10\%$ offset and upto 1-$\sigma$ (37.8-2.7=35.1) corresponds to $g\simeq 0.3$ (denoted by the red dashed line in \ref{SP4_Fig_M1b}). The red solid line corresponds to the centroid value of the angular diameter with $10\%$ offset. We however do not emphasize on this result because the spin corresponding to the red solid line is $a<0.5$. In this regard it is important to mention that numerical simulations indicate the spin of M87* to be either $a=0.5$ or $a=0.94$ \cite{Akiyama:2019fyp}. This result is obtained by fitting the numerical models with the observed jet power of M87*. This is the reason we draw vertical lines in \ref{SP4_Fig_M1} and \ref{SP4_Fig_M2} corresponding to $a=0.5$.

For completeness we mention that when $M\simeq 6.5\times 10^9 M_\odot$ is used to calculate the theoretical angular diameter $g\lesssim 0.02$ is required to explain the observed angular diameter within 1-$\sigma$ (blue dashed line in \ref{SP4_Fig_M2}. When $10\%$ offset in the observed angular diameter is considered then $g\lesssim 0.16$ is necessary (red solid line in \ref{SP4_Fig_M2}) to explain the observations. We do not emphasize much on this result because in this case the mass itself is derived from the observed shadow. With this mass therefore we cannot establish any constrains on $g$. We further note that a non-zero $g$ causes a decrease in the shadow size (\ref{SP4_Fig_S1_Vg}), and as a result a greater mass is required (as reported by the EHT collaboration) to explain the observed angular diameter (\ref{SP4_Phi}). Since a Kerr black hole casts a larger shadow than the Bardeen black hole, the observed image of M87* can be better explained by the Kerr scenario in \gr.

Another important and independent constraint on the non-linear electrodynamics charge parameter can be established with the help of the observable deviation from circularity $\Delta C$. According to the EHT collaboration the deviation from circularity for M87* is $\Delta C\lesssim 10\%$ \cite{Akiyama:2019cqa}. This leads to the constraint $g\lesssim 0.2$ from \ref{SP4_Fig_M3a}. which is deduced from the fact that for M87* $|a|\gtrsim 0.5$. The third observable $\Delta A$ cannot provide any bound on $g$ because the maximum value of $\Delta A$ from \ref{SP4_Fig_M3b} is 1.025 which is less than the EHT observation of $\Delta A\lesssim 4/3$ \cite{Akiyama:2019cqa}.

\subsection{Constrains on the magnetic monopole charge from the image of Sgr A*}
In this section we compare the theoretical angular diameter with the observed image of Sgr A*. According to the EHT collaboration the angular diameter of the emission ring of Sgr A* turns out to be $(51.8\pm2.3) \mu as$ \cite{EventHorizonTelescope:2022xnr,EventHorizonTelescope:2022vjs,EventHorizonTelescope:2022wok,EventHorizonTelescope:2022exc,EventHorizonTelescope:2022urf,EventHorizonTelescope:2022xqj}. However, the angular diameter of the shadow is $(48.7\pm 7)\mu a s$\cite{EventHorizonTelescope:2022xnr,EventHorizonTelescope:2022vjs,EventHorizonTelescope:2022wok,EventHorizonTelescope:2022exc,EventHorizonTelescope:2022urf,EventHorizonTelescope:2022xqj}. Since computation of the theoretical angular diameter depends on the mass and distance of the source, we first discuss the previously estimated distance and mass of Sgr A*. 

\begin{figure}[h!]%
    \centering
    \subfloat[Contours represent variation of angular diameter of Sgr A* with $a$ and $g$ for mass $M\simeq 3.951\times 10^{6}M_{\odot}$ and distance $D\simeq 7.935 \rm pc$]{{\includegraphics[width=7.5cm]{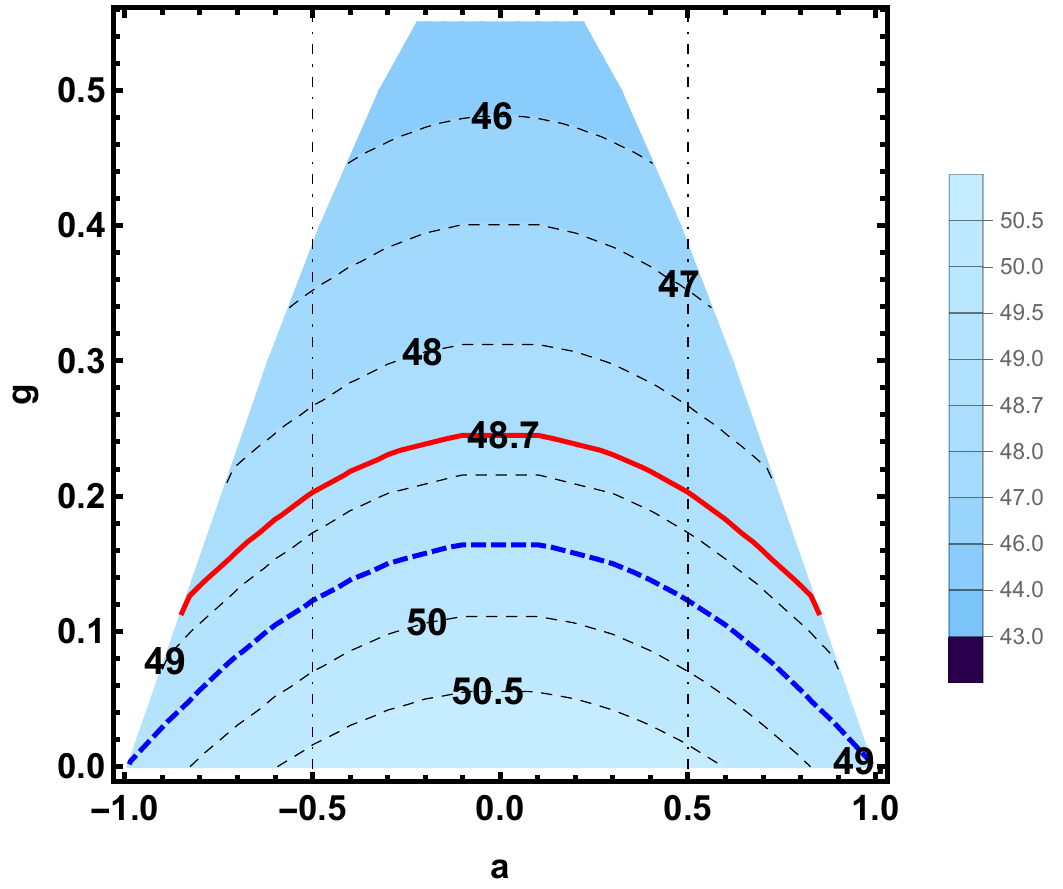} } \label{SP4_Fig_Sa}}
    \qquad
    \subfloat[Contours represent variation of angular diameter of Sgr A* with $a$ and $g$ for mass $M\simeq 3.975\times 10^{6}M_{\odot}$ and distance $D\simeq 7.959 \rm Kpc$]{{\includegraphics[width=7.5cm]{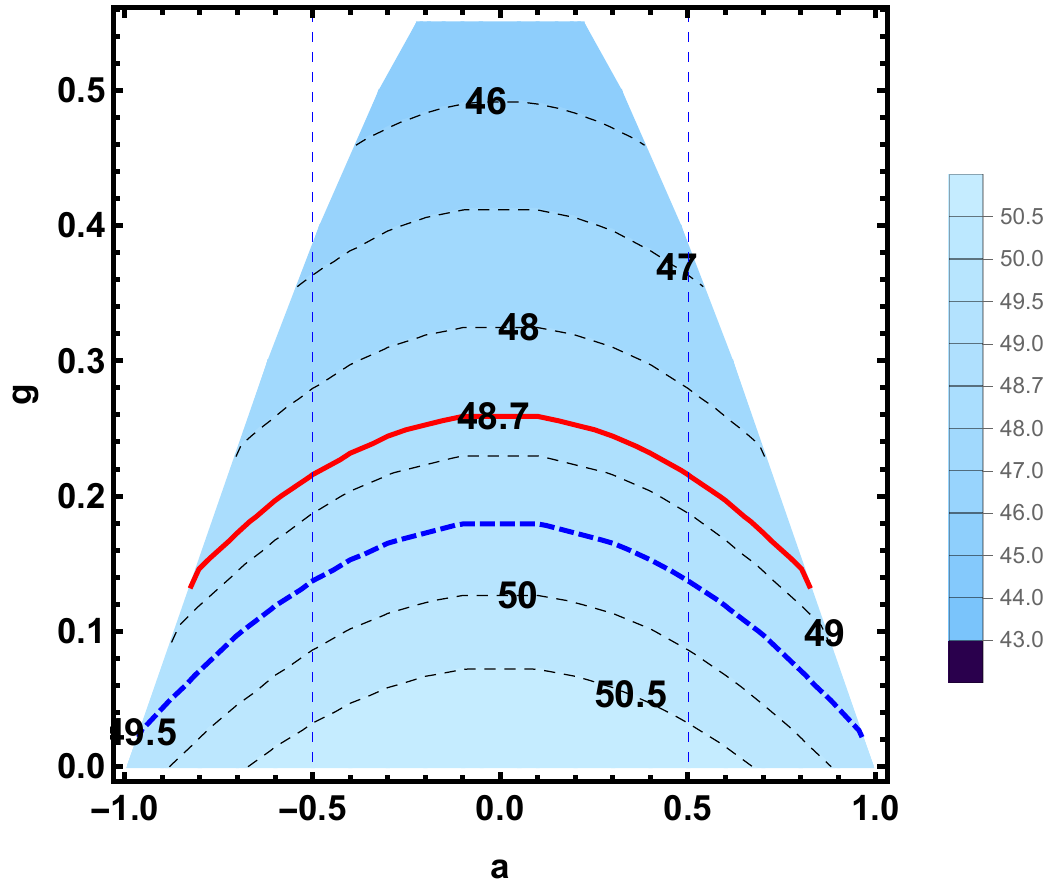} } \label{SP4_Fig_Sb}}
     \qquad
    \subfloat[Contours represent variation of angular diameter of Sgr A* with $a$ and $g$ for mass $M\simeq 4.261\times 10^{6}M_{\odot}$ and distance $D\simeq 8.277 \rm Kpc$]{{\includegraphics[width=7.5cm]{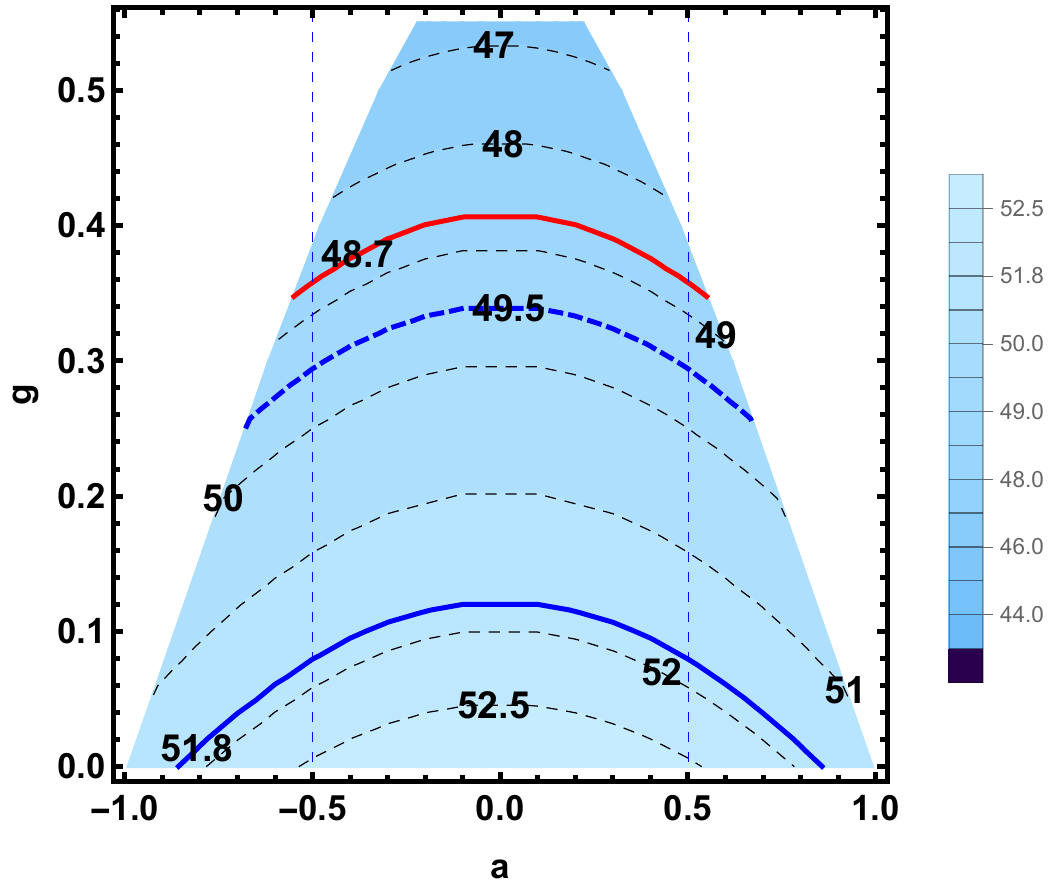} } \label{SP4_Fig_Sc}}
     \qquad
    \subfloat[Contours represent variation of angular diameter of Sgr A* with $a$ and $g$ for mass $M\simeq 4.297\times 10^{6}M_{\odot}$ and distance $D\simeq 8.2467 \rm Kpc$]{{\includegraphics[width=7.5cm]{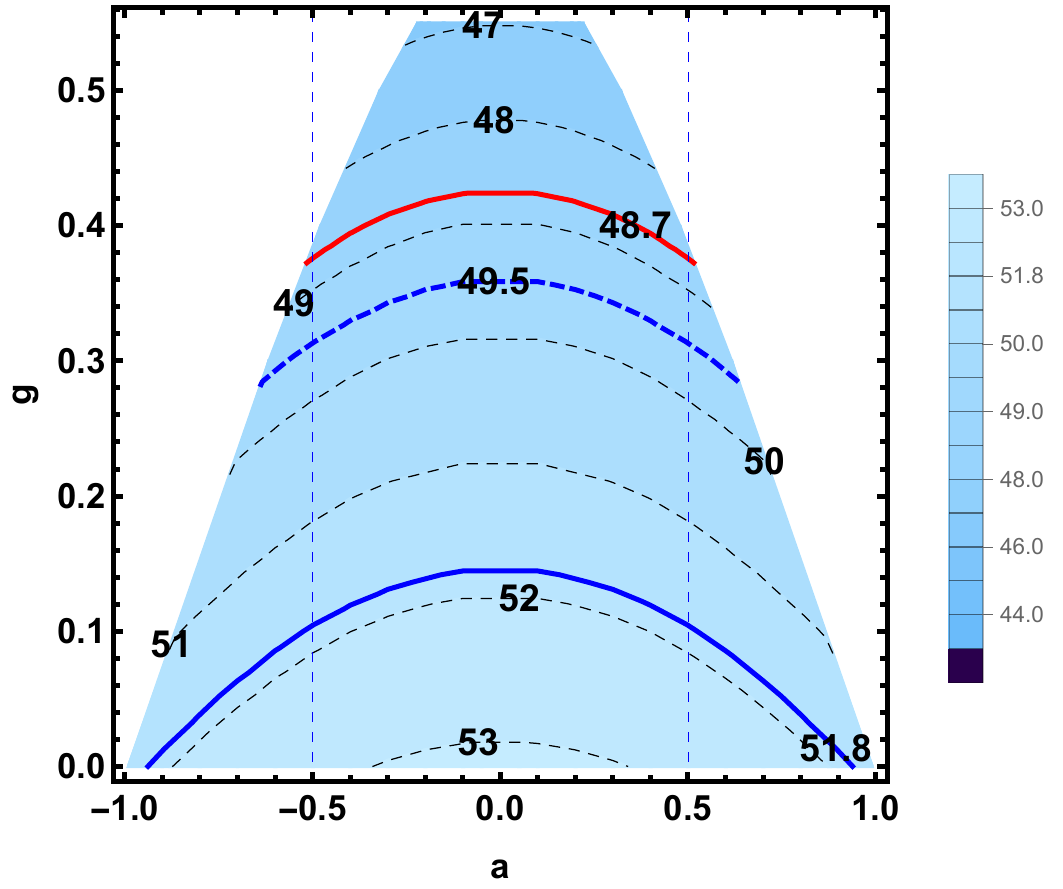} } \label{SP4_Fig_Sd}}
   \caption{Variation of angular diameter with $g$ and $a$ with various mass and distance estimations for Sgr A*. The inclination angle is taken to be $134^{\circ}$.}
    \label{SP4_Fig_S}
\end{figure}

The mass and distance of Sgr A* has been reported by several groups. According to the measurements of the Keck team Sgr A* has a distance of $D=(7959\pm59\pm32)\rm pc$ and mass $M=(3.975\pm 0.058\pm 0.026)\times 10^{6} M_{\odot}$, which is obtained by keeping the red-shift parameter free \cite{Do:2019txf}. The same group has reported distance $D=(7935\pm50)\rm pc$ and mass $M=(3.951\pm 0.047)\times 10^{6} M_{\odot}$, assuming the red-shift parameter to be unity \cite{Do:2019txf}. Estimates by the Gravity collaboration suggest the mass of Sgr A* to be $M=(4.261\pm0.012)\rm pc$ and distance $D=(8246.7\pm9.3)\rm pc$ \cite{GRAVITY:2021xju,GRAVITY:2020gka}. By taking into account the optical aberrations the Gravity collaboration further constrained the BH mass $M=(4.297\pm0.012\pm0.040)\times 10^{6} M_{\odot}$ and the distance $D=8277\pm9\pm33 \rm pc$.
For obtaining the theoretical angular diameter of Sgr A* one needs the inclination angle of the BH, which in our case is $i=134^{\circ}$ \cite{2019A&A...625L..10G}. Using this inclination angle we plot the variation of the angular diameter with the charge parameter $g$ and the rotation paameter $a$ in \ref{SP4_Fig_S} for various combinations of masses and distances as discussed above.

\begin{figure}[h!]%
    \centering
    \subfloat[Above figure shows the variation of the deviation from circularity $\Delta C$ with $g$ and $a$ for Sgr A* with an inclination angle $134^{\circ}$]{{\includegraphics[width=7.5cm]{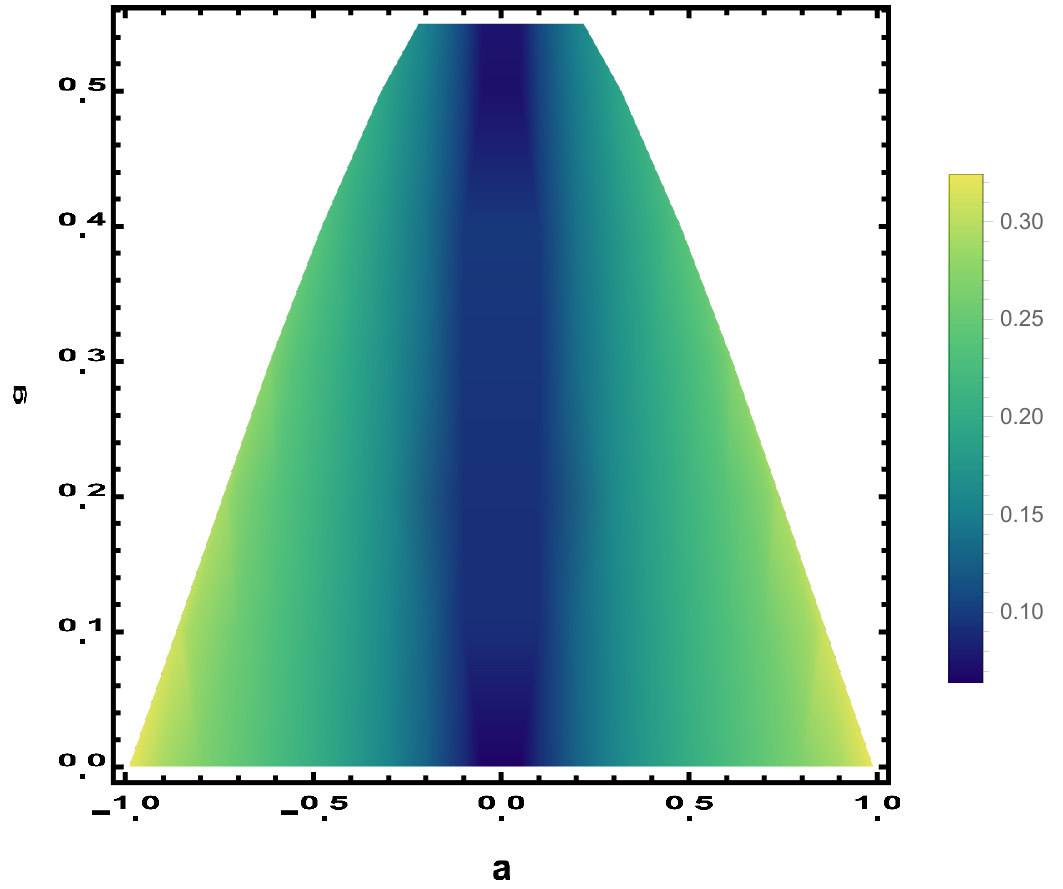} } \label{SP4_Fig_S1a}}
    \qquad
    \subfloat[Above figure shows the variation of the axis ratio $\Delta A$ with $g$ and $a$ for Sgr A* with an inclination angle $134^{\circ}$]{{\includegraphics[width=7.5cm]{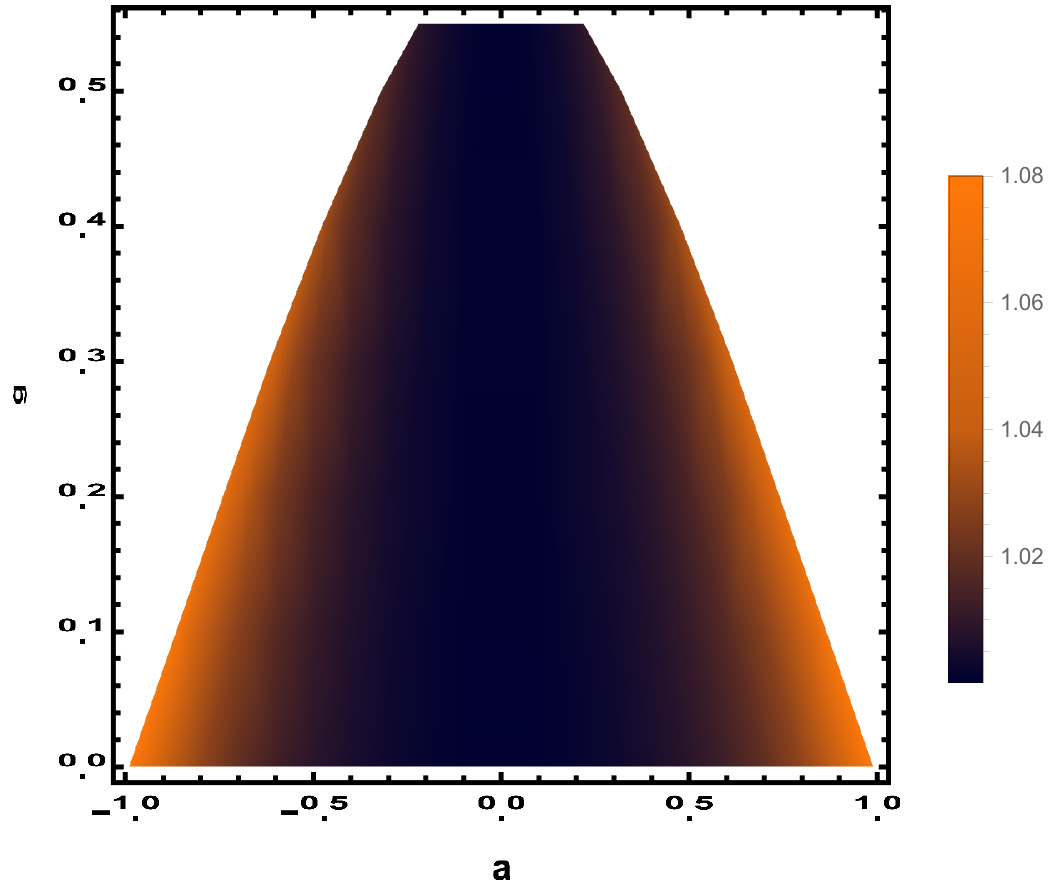} } \label{SP4_Fig_S1b}}
    \caption{We have plotted the variation of axis ratio $\Delta A$ and deviation from circularity $\Delta C$ in the $(g-a)$ plane for Sgr A* with an inclination angle $134^{\circ}$ }
    \label{SP4_Fig_S1}%
\end{figure}

When mass and distance reported by the Keck team (\ref{SP4_Fig_Sa} and \ref{SP4_Fig_Sb}) is used, $0.15\lesssim g\lesssim 0.2$ is required to reproduce the centroid value of the observed shadow diameter ($48.7\mu as$, denoted by the red solid line in \ref{SP4_Fig_Sa} and \ref{SP4_Fig_Sb}). We further note that $0.04\lesssim g \lesssim 0.1$ can explain the observed image diameter within 1-$\sigma$ (denoted by the blue dashed line in \ref{SP4_Fig_Sa} and \ref{SP4_Fig_Sb}). Like M87*, for Sgr A* as well numerical models that best explain the observed emission ring correspond to $|a|=0.5$ and $|a|=0.94$ \cite{EventHorizonTelescope:2022urf,EventHorizonTelescope:2022xqj}. This is consistent with the spin estimates of Sgr A* from its radio spectrum \cite{Reynolds:2013rva,Moscibrodzka:2009gw,Shcherbakov:2010ki} which reports $a\sim 0.9$ \cite{Moscibrodzka:2009gw} and $a\sim0.5$ \cite{Shcherbakov:2010ki}. Therefore in \ref{SP4_Fig_S} the vertical dashed lines are drawn for $|a|=0.5$. 

When mass and distance reported by the Gravity collaboration is used $0.34\lesssim g \lesssim0.36$ is required to explain the observed shadow diameter (which is $48.7\mu as$ denoted by the red solid line in \ref{SP4_Fig_Sc} and \ref{SP4_Fig_Sd}). However, when one intends to explain the diameter of the centroid value of the emission ring (51.8 $\mu as$ denoted by the blue solid line in \ref{SP4_Fig_Sc} and \ref{SP4_Fig_Sd} ) $g$ can be as high as $0.06$ for $M=4.261\times 10^6 M_\odot$ and $D=8.277$ kpc and $0.08$ for $M=4.297\times 10^6 M_\odot$ and $D=8.2467$ kpc. When the 1-$\sigma$ interval in the emission ring is considered a higher value of the magnetic monopole charge parameter $0.25\lesssim g\lesssim 0.3$ is allowed. This contour is represented by the blue dashed line in \ref{SP4_Fig_Sc} and \ref{SP4_Fig_Sd}. From the above discussion it is clear that a small but non-zero value of $g$ can explain the observed shadow of Sgr A* better than the Kerr scenario. This is attributed to the fact that increase in $g$ leads to a decrease in the shadow size (\ref{SP4_Fig_S1_Vg}) 

The Event Horizon Telescope collaboration does not impose any constrains on $\Delta C$ and $\Delta A$ for the image of Sgr A*. We however present plots corresponding to the two aforesaid observables assuming the inclination angle $i=134^\circ$ which can be used for future work once these bounds are reported by the EHT collaboration. This is illustrated in 
\ref{SP4_Fig_S1} for completeness.

\section{Conclusion}
\label{S5}
In this work we investigate the role of the magnetic monopole charge arising in the Bardeen black hole scenario in explaining the observed shadow of M87* and Sgr A*. Bardeen black holes are interesting as they have a regular core and hence can potentially avoid the $r=0$ curvature singularity arising in \gr. In the absence of a well accepted quantum theory of gravity regular black holes provide a classical route to address the singularity issue in GR thereby making the theory finite at all length scales. Such black holes generally arise in Einstein gravity coupled to non-linear electrodynamics which becomes relevant when the electromagnetic field is very strong. Therefore, it is worthwhile to explore the signatures of regular black holes in astrophysical observations. With two successive release of black hole images by the EHT collaboration, the scope to test the nature of srong gravity has enhanced considerably.

We study the nature of the shadow cast by Bardeen black holes and explore the effect of the magnetic monopole charge parameter on the shape and size of the shadow. Such a study reveals that with an increase in the inclination angle and the black hole spin the shadow becomes more dented. Further, an increase in the magnetic monopole charge parameter leads to a decrease in the shadow size. This feature plays a pivotal role in explaining the observed angular diameter of the image of Sgr A* and M87*. 

For M87*, when the theoretical angular diameter is calculated with mass estimated from gas dynamics measurements, the observed shadow diameter of $42\pm 3 \mu as$ cannot be reproduced. When mass reported from stellar dynamics measurements is used, the observed shadow diameter can be explained within 1-$\sigma$ only when the maximum offset of 10\% is considered in the image diameter. Since a non-zero magnetic monopole charge $g$ shrinks the shadow, the centroid value of the observed shadow of M87* can be better explained by the Kerr scenario. 

For Sgr A*, the mass and distance is very well constrained. When the distance and mass estimated by the Keck team is used, $0.15\lesssim g \lesssim 0.2$ is required to reproduce the centroid value of the observed angular diameter of Sgr A*. Similarly, $0.34 \lesssim g\lesssim 0.36$ explains the central value of the shadow diameter ($48.7 \mu as$) of Sgr A* when mass and distance reported by the Gravity collaboration is used. Therefore, for Sgr A* the Bardeen black hole scenario better explains its observed shadow. 

We have previously studied the prospect of Bardeen black holes in explaining the optical observations of quasars \cite{Banerjee:2021nza} and found that quasar optical data favors the Kerr scenario. A similar conclusion has been reached when we studied the possibility of Bardeen black holes in explaining the observed quasi-periodic oscillations \cite{Banerjee:2022ffu}. Finally in the present work we note that the shadow of M87* can be better explained by the Kerr scenario while the image of Sgr A* favors the Bardeen black hole framework. Thus, although rare, black holes do exhibit signatures of magnetic monopole charge in certain cases. 
%It is important to note that  analysis of the continuum spectrum, the quasi-periodic oscillations (QPOs) and the black hole shadow are done for different observational samples. 
The study of the nature of spacetime with respect to different black holes can be better accomplished in the near future with the availability of more images of black holes with improved resolution.

\section*{Acknowledgements}

\bibliography{Black_Hole_Shadow,Black_Hole_Shadow2,IB,Brane,KN-ED,regularBh,regularBh2,bardeen,SgrA,QPO}
\bibliographystyle{./utphys1}

%\bibliographystyle{utphys1}
%\bibliography{x.bib}
%\bibliographystyle{plain}
%\bibliographystyle{JHEP}
%\bibliographystyle{plainnat}
\end{document}